%% file: main.tex
\documentclass[11pt, a4paper]{article}

\input{src/preamble}

\title{A Comparison of Surrogate Constitutive Models for Viscoplastic Creep Simulation of HT-9 Steel}

\author{
  Pieterjan Robbe\thanks{Sandia National Laboratories, Livermore, CA 94550, USA} \and
  Andre Ruybalid\thanks{Los Alamos National Laboratory, Los Alamos, NM 87545, USA} \and
  Arun Hegde\footnotemark[1] \and
  Christophe Bonneville\footnotemark[1] \and
  Habib N.\ Najm\footnotemark[1] \and
  Laurent Capolungo\footnotemark[2] \and
  Cosmin Safta\footnotemark[1]
}

\date{} % leave empty or add submission date if you like

\begin{document}

\maketitle

\begin{abstract}
Mechanistic microstructure-informed constitutive models for the mechanical response of polycrystals are a cornerstone of computational materials science. However, as these models become increasingly more complex -- often involving coupled differential equations describing the effect of specific deformation modes -- their associated computational costs can become prohibitive, particularly in optimization or uncertainty quantification tasks that require numerous model evaluations. To address this challenge, surrogate constitutive models that balance accuracy and computational efficiency are highly desirable. Data-driven surrogate models, that learn the constitutive relation directly from data, have emerged as a promising solution. In this work, we develop two local surrogate models for the viscoplastic response of a steel: a piecewise response surface method and a mixture of experts model. These surrogates are designed to adapt to complex material behavior, which may vary with material parameters or operating conditions. The surrogate constitutive models are applied to creep simulations of \texorpdfstring{\HT{9}}{HT-9} steel, an alloy of considerable interest to the nuclear energy sector due to its high tolerance to radiation damage, using training data generated from viscoplastic self-consistent (VPSC) simulations. We define a set of test metrics to numerically assess the accuracy of our surrogate models for predicting viscoplastic material behavior, and show that the mixture of experts model outperforms the piecewise response surface method in terms of accuracy.
\end{abstract}

\hypersetup{
    linkcolor=RubineRed,
    urlcolor=OliveGreen,
    citecolor=SeaGreen
}

\input{src/introduction}

\input{src/methods}

\input{src/results}

\input{src/discussion}

\input{src/conclusion}

\input{src/acknowledgements}

\bibliographystyle{abbrv}
\bibliography{src/references.bib}

\appendix
\input{src/appendix}

\end{document}

%% file: src/preamble.tex
\usepackage{ifthen}
\newboolean{debugmode} %
\setboolean{debugmode}{false} %

\usepackage[margin=1in]{geometry}
\usepackage[T1]{fontenc}
\usepackage{amsfonts}
\usepackage{amsmath}
\usepackage{mathtools}
\usepackage{mathrsfs}
\usepackage{color, soul}
\usepackage{graphicx}
	\graphicspath{{plt/}, {ext/}} %
\usepackage[dvipsnames]{xcolor}
\usepackage{tikz}
	\usetikzlibrary{external, positioning, arrows.meta, calc, decorations.markings, shapes.geometric, patterns}
\usepackage{pgfplots}
	\pgfplotsset{compat=newest}
	\usepgfplotslibrary{colorbrewer, colormaps, statistics, groupplots, fillbetween}
	\newlength{\figurewidth}
	\newlength{\figureheight}
\usepackage{pgfplotstable}
\usepackage{booktabs}
\usepackage[version=4]{mhchem} %
\usepackage[
  colorlinks = true, %
  urlcolor   = blue, %
  linkcolor  = blue, %
  citecolor  = blue, %
  ]{hyperref}
\PassOptionsToPackage{hyphens}{url}\usepackage{hyperref}
\usepackage{xurl}
\usepackage[capitalise, noabbrev, nameinlink]{cleveref}
    \crefname{equation}{equation}{equations}
\usepackage{siunitx}
\usepackage{dsfont} %
\usepackage{multirow}
\usepackage[normalem]{ulem} %
\usepackage{tcolorbox} %
    \tcbuselibrary{most}
\usepackage[clock]{ifsym}
\usepackage{nicematrix}
\NiceMatrixOptions{%
    custom-line={%
        letter=I,
        tikz={dashed, line cap=round, semithick, dash phase=3pt},
        total-width=\pgflinewidth
    }
}

\colorlet{Color1}{RoyalBlue}
\colorlet{Color2}{Red}
\colorlet{Color3}{ForestGreen}
\colorlet{Color4}{Dandelion}

\input{src/mathdefs}

\newcommand{\etal}{et al.}

\newcommand{\dd}{\mathrm{d}}

\newcommand{\vmJ}[1]{\sigma_\text{vm}}
\newcommand{\T}{T}
\newcommand{\evm}{\varepsilon_\text{vm}}
\newcommand{\rhoc}{\rho_\text{cell}}
\newcommand{\rhow}{\rho_\text{wall}}
\newcommand{\flux}{\dot{\phi}}

\newcommand{\evmdot}{{\dot{\varepsilon}_\text{vm}}}
\newcommand{\rhocdot}{{\dot{\rho}_\text{cell}}}
\newcommand{\rhowdot}{{\dot{\rho}_\text{wall}}}

\newcommand{\HT}[1]{\text{HT-9}}

\DeclareSIUnit\dpa{dpa}

\tikzset{%
    >={Stealth[length=2mm, width=1.75mm]}, %
    default line/.style={%
        semithick,
        line cap=round,
    },
    default dashed line/.style={%
        default line,
        dashed,
    },
    default marker line/.style={%
        default line,
        mark=*,
        mark size=0.75pt,
    },
}

\pgfkeys{%
	/pgf/number format/.cd,1000 sep={}
}

\pgfplotsset{%
    default axis/.style={%
        width=\figurewidth,
        height=\figureheight,
        major tick length={2pt},
        minor tick length={2pt},
        every tick/.style={black, line cap=round},
        ticklabel style={font=\scriptsize},
        label style={font=\small},
        legend style={%
            draw=none,
            font=\scriptsize,
            at={(1.03, 1)},
            anchor=north west,
            fill=none,
            legend cell align=left
        },
        cycle list/Set1,
    },
    default error bar/.style={%
        error bars/.cd,
        y dir=both,
        y explicit,
    },
    default scatter/.style={%
        scatter,
        only marks,
        scatter src=explicit,
        mark size=1.75pt,
        mark options={%
            line width=0.1pt
        }
    },
        colormap/YlOrRd,
        colormap/RdBu,
}

%% file: src/mathdefs.tex
\newcommand{\bsb}{{\boldsymbol{b}}}

\newcommand{\bsg}{{\boldsymbol{g}}}

\newcommand{\bsw}{{\boldsymbol{w}}}
\newcommand{\bsx}{{\boldsymbol{x}}}
\newcommand{\bsy}{{\boldsymbol{y}}}
\newcommand{\bsz}{{\boldsymbol{z}}}

\newcommand{\bsalpha}{{\boldsymbol{\alpha}}}

\newcommand{\bsphi}{{\boldsymbol{\phi}}}

\newcommand{\bsomega}{{\boldsymbol{\omega}}}

\newcommand{\bsPsi}{{\boldsymbol{\Psi}}}

\newcommand{\bbR}{{\mathbb{R}}}

\DeclareSymbolFont{bbold}{U}{bbold}{m}{n}
\DeclareSymbolFontAlphabet{\mathbbold}{bbold}

\newcommand{\calE}{{\mathcal{E}}}
\newcommand{\calF}{{\mathcal{F}}}

\newcommand{\calL}{{\mathcal{L}}}
\newcommand{\calM}{{\mathcal{M}}}

\newcommand{\calT}{{\mathcal{T}}}

\providecommand{\sign}{\operatorname*{sign}}

\providecommand{\argmin}{\operatorname*{argmin}}

%% file: src/introduction.tex
\section{Introduction}\label{sec:introduction}

The development of models that predict the mechanical response of complex metals as a function of their microstructure is critical for material design, material selection, and qualification. In this context, material models must be valid over a wide spectrum of thermomechanical constraints and be able to extrapolate under conditions that cannot be easily tested experimentally. To this end, the computational materials community has, over the past several decades proposed to use multiscale modeling frameworks, seamless or not, to derive such types of models~\cite{horstemeyer2009}. The ensuing mechanistic and microstructure informed models keep track of the relative and absolute contributions of the multitude of deformation modes and physical processes during plastic deformation whilst relating those to the dynamics of the microstructure evolutions. 

From a numerical point of view, mechanistic models for the response of metals are complex as they rely on (i) the use of constitutive laws (e.g. crystal plasticity, strain gradient plasticity) which relate both the elastic and plastic strains at a material point to its stress~\cite{asaro1983,gao1999}, and, (ii) on homogenization models relying either on mean-field Eshelbian schemes or on full-field methods (e.g.\,Fast Fourier based mechanical solvers, Finite Element Solvers) to predict the overall response of a representative volume of the polycrystalline assembly~\cite{lebensohn2011}. Naturally, the refinement of these multiscale models (i.e.\,from the single crystal length scale to the polycrystal scale) also leads to an increase in computational complexity. For example, when studying the viscoplastic behavior of polycrystalline materials, such as metals and alloys, the Viscoplastic Self-Consistent (VPSC) mean field homogenization model from Lebensohn \etal\,\cite{lebensohn1993, lebensohn2004} combined with advanced constitutive models~\cite{wang2016,wen2020} that track internal state variables results in a set of coupled ordinary differential equations that describe the evolution of stress, strain, and internal state variables for individual grains within a polycrystalline material.

To realize their full prospects, these polycrystal models can be either integrated in finite element solvers, thereby allowing to simulate the response of components with a sensitivity to the underlying microstructure of the constituents~\cite{knezevic2013, patra2017} or used as part of an uncertainty quantification exercise~\cite{khadyko2018}. In both such applications, the numerical cost of mechanistic polycrystal models can be computationally prohibitive, although we note that some developments have been made to this end~\cite{ruybalid2024}.

To address this challenge, surrogate constitutive models that trade accuracy for simulation cost have become paramount in computational workflows \cite{dornheim2024}. These surrogate models serve as simplified approximations of traditional constitutive relations, by balancing computational efficiency with the fidelity of material response predictions. Surrogate constitutive models can be constructed using various methods, including response surface methods~\cite{daoud2014}, polynomial chaos expansions~\cite{tallman2021, he2022}, and Gaussian processes~\cite{frankel2020, rocha2021, fuhg2022}. These techniques leverage existing data from high-fidelity simulations or experimental results to create a mapping between input parameters, such as stress, strain, and temperature, and the corresponding material responses. The resulting surrogate models can be implemented within a larger computational framework, allowing for rapid evaluations of material behavior under varying conditions without the need for repeated evaluations of the original, computationally demanding, constitutive models~\cite{segurado2012, roters2010}. Hence, they allow for a computationally efficient exploration of design spaces and the optimization of material performance in practical engineering applications~\cite{forrester2008}.

Data-driven approaches for constitutive models have been proposed several decades ago, and have resurfaced in recent years due to the growth in machine learning methods and tools. Ghaboussi \etal\,\cite{ghaboussi1991} explored the use of neural networks (NNs) to model stress-strain relationships in composites. Furukawa and Yagawa~\cite{furukawa1998} proposed a neural constitutive model for viscoplasticity. In this work, a NN was trained to predict rates associated with the viscoplastic strain and other internal variables. Jung and Ghaboussi~\cite{jung2006} implemented a neural constitutive model in the finite element analysis of a concrete beam undergoing a creep test. More recently, NN-based constitutive relations have been applied to study heterogeneous elasticity~\cite{ye2019, ye2021}, rate-independent plasticity~\cite{al2006, yang2020, zhang2022}, rate-dependent plasticity~\cite{li2019, pandya2020, wen2021}, and crystal plasticity~\cite{ali2019}. Various NN architectures have been considered in these works, including graph neural networks~\cite{storm2024}, neural operators~\cite{jafarzadeh2024} and physics-informed neural networks~\cite{linka2021, dornheim2024}. These NN-based constitutive models offer a significant advantage over traditional surrogate constitutive models, because they have the ability to represent arbitrary complex material behavior. For a more detailed overview on machine learning-based constitutive models, we refer to recent surveys~\cite{dornheim2024,fuhg2024,rosenkranz2023,herrmann2024}.

The data-driven surrogate constitutive models in these previous works have two major drawbacks:
\begin{enumerate}
    \item They are derived from relatively simple constitutive laws, and are not necessarily designed to operate over a wide range of mechanisms and operating conditions.
    \item While several methods have been proposed to derive these models, it is difficult to assess which method yields the best trade-off in terms of accuracy and computational cost.
\end{enumerate}
In this work, we develop two local constitutive model surrogates~\cite{tallman2021}, where the input parameter space is divided into smaller, localized regions, each of which is associated with a simplified constitutive model that captures the essential behavior of the material within that specific region:
\begin{itemize}
    \item A piecewise \emph{Response Surface Methodology} (RSM) surrogate~\cite{khuri2010} that forms an explicit tiling of the input space. Within each tile, the constitutive model response is approximated by a low-order polynomial, enabling explicit interpolation of constitutive responses across the input space.
    \item A \emph{Mixture of Experts} (MoE) surrogate~\cite{masoudnia2014, shazeer2017} where multiple experts are used to automatically divide the input space into homogeneous regions. Each expert is modeled as a data-driven constitutive surrogate, and the information of each expert is combined using an expert weighting function.
\end{itemize}
These surrogate constitutive models in part address the aforementioned drawbacks: we will apply them in a complex polycrystal model setting prototypical of advanced viscous laws (cf.\,drawback \#1); we systematically compare the advantages and limitations of each model (cf.\,drawback \#2). %

The RSM was originally proposed by Box and Wilson~\cite{box1992}. The main idea of this method is to construct an efficient surrogate through the careful selection of training data (experimental design) and using least-squares regression to find the coefficients of a low-order polynomial fit through the data. In practical applications, linear or quadratic polynomials are most common. RSMs have been used as surrogate constitutive models in, among others, Shen \etal\,\cite{shen2021} and Daoud \etal\,\cite{daoud2014}.

Adaptive MoEs were first proposed by Jacobs \etal\,\cite{jacobs1991} in the context of Gaussian mixture models. The main idea is to dynamically partition the input space to create a sequence of smaller model fitting tasks that are easier to accomplish~\cite{yuksel2012}. MoEs gained renewed interest in the context of deep learning in Eigen \etal~\cite{eigen2013}. In this work, each expert is a linear NN with ReLU activation functions, and the weighting function is a simple NN followed by a softmax layer. Morand and Helm~\cite{morand2019} have used MoEs for surrogate constitutive models in the context of elasto-plastic deformation of metallic materials during tensile tests.

We apply our novel surrogate constitutive modeling strategies in the context of viscoplasticity in \HT9{} steel. \HT9{} (\ce{Fe-12Cr-1Mo}) is a high-chromium ferritic/martensitic stainless steel alloy that is primarily utilized as a cladding material for nuclear fuel rods in fast neutron reactors~\cite{wen2020, wen2017, wen2021}. Its combination of high strength, corrosion resistance, radiation tolerance and thermal stability makes a promising material system for next generation nuclear reactors. The underlying mechanistic model was introduced and implemented within a VPSC framework in Wen \etal\,\cite{wen2020}. This model was shown to be able to quantify multiple physical mechanisms, including dislocation glide, dislocation climb, vacancy-mediated diffusional creep, and, albeit in a simplified fashion, irradiation effects in \HT{9} alloys. The constitutive model describes the evolution of the equivalent plastic strain rate, and uses a dislocation density evolution model~\cite{wen2017} to track dislocations inside sub-grains (``cells'') and at sub-grain boundaries (``cell walls''), as a function of imposed von Mises stress, temperature, irradiation dose rate, current dislocation content and accumulated plastic strain during creep. In this setting, surrogate models that track these dislocation densitities as internal state variables must be able to capture a wide range of temperatures and stresses, which presents a significant challenge for conventional global surrogate constitutive models that are typically trained over the entire input space and may struggle with local accuracy in regions of complex behavior. To address this, the proposed methods adopt a localized modeling strategy, which enables the surrogate constitutive models to better adapt to complex material behavior, where variations in material parameters and operating conditions can lead to sharp changes in the material response.

The original contributions of this work are as follows:
\begin{itemize}
    \item We develop two novel local surrogate constitutive models for viscoplasticity that use a partitioning of the input parameter space (either explicitly through the piecewise RSM or implicitly through MoEs) to improve point-wise accuracy of the constitutive surrogate.
    \item We apply these new surrogates to predict the behavior of \HT9 steel under creep loading, and compare the accuracy of the predictions against reference VPSC data using well-defined accuracy metrics.
    \item We compare the advantages and limitations of each method for modeling viscoplastic material behavior.
\end{itemize}

The remainder of this paper is organized as follows. In~\cref{sec:methods}, we outline our proposed methodology in more detail. We discuss surrogate constitutive models, and introduce two novel local surrogate models. Next, in~\cref{sec:results}, we present the results obtained by applying the two data-driven constitutive models to a reference creep loading example involving \HT9{} alloys. In \cref{sec:discussion}, we review these results in more detail, and discuss conclusions and future work.

%% file: src/methods.tex
\section{Methods}
\label{sec:methods}

In this section, we outline our methodology for surrogate constitutive model construction. We start by discussing our model for viscoplasticity in \HT9 steel and discuss our strategy for constructing data-driven surrogate constitutive models. We then describe in more detail the two approaches for local surrogate construction we consider in this work. Here, we discuss these surrogates to model a general input-output mapping only, and important details (including database generation and input and output transformations) are deferred to \cref{subsec:database_generation,subsec:input_output_transformations}, respectively.

\subsection{Viscoplasticity in \texorpdfstring{\HT{9}}{HT-9} steel}
\label{sec:viscoplasticity_in_HT9_steel}
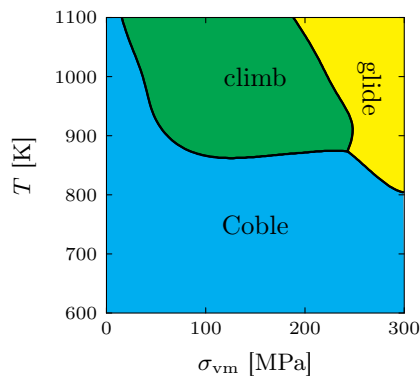
\begin{figure}[ht]
    \centering
    \input{fig/creep_models}
    \caption{Dominant creep mechanisms in \HT{9} steel as a function of stress $\vmJ{2}$ and temperature $T$.}
    \label{fig:creep_models}
\end{figure}

The viscoplastic deformation behavior of ferritic/martensitic \HT{9} steel arises from several competing microscopic mechanisms, including dislocation glide, dislocation climb, and diffusion-controlled creep. These mechanisms operate across a broad range of stress and temperature conditions, as illustrated in Figure~\ref{fig:creep_models}, and their relative contributions depend sensitively on the microstructure and external loading conditions, as well as irradiation effects (relevant for nuclear cladding materials like \HT{9}). The total viscoplastic strain rate can be expressed as the additive contribution of these modes:
\begin{equation}\label{eq:total_creep_rate}
    \dot\varepsilon_{ij} = \dot{\varepsilon}_{ij}^\text{glide} + \dot{\varepsilon}_{ij}^\text{climb} + \dot{\varepsilon}_{ij}^\text{diff},
\end{equation}
where each component reflects a distinct physical mechanism that governs deformation. As also shown in Figure~\ref{fig:creep_models}, the relative and absolute contributions of these mechanisms vary significantly with temperature, necessitating modeling frameworks that explicitly account for temperature dependence. 
In this work, we utilize an existing mechanistic model~\cite{wen2020} embedded in a viscoplastic self-consistent (VPSC) framework~\cite{lebensohn2004}. For completeness, a brief overview of the VPSC-based model is provided here and the reader is referred to the original publication by Wen \etal~\cite{wen2020} for more detailed information. The diffusional strain component is based on the Coble creep formalism~\cite{coble1963}, which contributes to plastic deformation in polycrystals via migration of point defects along grain boundaries, and dominates the viscoplastic deformation at lower stress and elevated temperature regimes. The two dislocation-based strain rate contributions from glide and climb are computed over all active slip systems $s$ in a given grain. Each shear or climb event contributes to the macroscopic viscoplastic strain as
\begin{equation}\label{eq:Orowan}
    \dot{\varepsilon}_{ij}^\text{glide} = \sum_s m_{ij}^s \dot\gamma^s, \qquad \dot{\varepsilon}_{ij}^\text{climb} = \sum_s c_{ij}^s \dot\beta^s, 
\end{equation}
where $m^s_{ij}$ represents the symmetric Schmid tensor, and $c^s_{ij}$ represents the ``climb'' tensor~\cite{lebensohn2010}, and where, $\dot\gamma^s$ and $\dot\beta^s$ denote the mean shear and climb rates in slip system $s$, respectively. The quantity $\dot\gamma^s$ characterizes the rate of plastic shear deformation due to dislocation glide in system $s$, and is the primary mode of plasticity at higher stress values. In contrast, $\dot\beta^s$ represents the mean climb rate of dislocations in system $s$ and becomes increasingly important at medium stress levels and higher temperatures, or under irradiated conditions where point defect mobility is enhanced. Climb allows dislocations to move out of their glide planes, enabling plastic flow even in directions not aligned with the primary slip systems, and plays a crucial role in creep and irradiation-assisted deformation mechanisms. These shear and climb rates are functions of dislocation densities and their mobilities, and thus encode microstructure state variables such as dislocation density. The shear and climb rates per slip system in body-centered cubic (BCC) \HT{9} are dependent on the microstructure state variables as
\begin{equation}
    \dot{\gamma}^s = f_\gamma(T, \mathbf{\sigma}, \rhoc, \rhow) \quad\text{and}\quad
    \dot{\beta}^s = f_\beta(T, \mathbf{\sigma}, \rhoc, \rhow, \flux),
\end{equation}
where $\rhoc$ and $\rhow$ represent the dislocation density in sub-grains within a single crystal (here denoted as ``cell'' dislocations) and at sub-grain boundaries within a single crystal (denoted as ``cell-wall'' dislocations). Additionally, the dislocation climb rate depends on the irradiation dose rate $\flux$, which quantifies the rate of generation of irradiation-induced point defects and is typically expressed in units of displacements per atom per second (\unit{\dpa\per\second}). A higher irradiation dose rate increases the steady-state concentration of point defects, thereby enhancing diffusional mechanisms that facilitate dislocation climb. While this can promote creep deformation at high doses or temperatures, the net effect during service is often a competition between irradiation defect-induced hardening -- which is not considered in the present model -- and climb-assisted recovery mechanisms. Moreover, the constitutive model explicitly tracks dislocation density quantities (both cell and cell-wall dislocations), using a dislocation density evolution model~\cite{wen2017}, making these microstructure quantities available for further modeling efforts.

As outlined in \cref{sec:introduction}, the complexity and computational cost of running high-fidelity VPSC simulations over wide parametric spaces motivate the use of surrogate models. We generate a comprehensive synthetic database of creep responses using the VPSC model over a range of stress, temperature, initial dislocation states, and irradiation dose rates. This database serves as the foundation for training data-driven surrogate models. Details regarding this database generation will be discussed in \cref{subsec:database_generation}. To make the surrogates computationally efficient while preserving essential physics, we reduce the dimensionality of the input and output spaces.

In this context, we assume approximate isotropy in the plastic response of \HT{9} due to its BCC crystal structure and the use of high grain counts in the simulations, which effectively averages out crystallographic anisotropy. This assumption allows us to use scalar effective measures of stress and strain rate, derived from the von Mises formulation. In particular, we define the effective stress as
\begin{equation}
    \vmJ{2} = \sqrt{\frac{3}{2} s_{ij} s_{ij}},
    \label{eq:effective_stress}
\end{equation}
with $s_{ij}$ the deviatoric Cauchy stress tensor, and approximate the viscoplastic strain rate direction via the Prandtl--Reuss flow rule
\begin{equation}
    \dot\varepsilon_{ij} = \left(\frac{3}{2} \evmdot\right) \frac{\partial s_{ij}}{\partial \vmJ{2}},
    \label{eq:normality_rule}
\end{equation}
where $\evmdot$ is the equivalent strain rate and $s_{ij}$ is the deviatoric stress tensor. Importantly, this simplification is applied only in the surrogate model to enable tractable learning of the strain-rate evolution from reduced input variables.

The surrogate thus takes as inputs: effective strain $\evm$, effective stress $\vmJ{2}$, temperature $\T$, irradiation dose rate $\flux$, and average dislocation densities $\rhoc$ and $\rhow$. It predicts the instantaneous viscoplastic strain rate $\evmdot$ and the evolution of dislocation densities $\rhocdot$ and $\rhowdot$. While this dimensionality reduction omits full tensorial fidelity, it preserves the essential features needed for effective prediction of \HT{9}'s thermomechanical response in practical settings.

In particular, we look for a mapping
\begin{equation}
    \calM : (\evm, \vmJ{2}, \T, \flux, \rhoc, \rhow) \mapsto (\evmdot, \rhocdot, \rhowdot).
\end{equation}
Note that this formulation is model agnostic, i.e., no explicit functional relation between inputs and outputs is assumed. Instead, this relationship will be learned from data.

Assuming a creep loading scenario, the material evolution can be uniquely specified by providing initial conditions for the strain and dislocation densities, as well as the control input $\vmJ{2}(t) = 0$ for all time $t > 0$, i.e.,
\begin{equation}
    \begin{cases}
        \begin{aligned}
            \evm(0)   &= 0, &&\\
            \rhoc(0)  &= \rho_{\text{cell}, 0},      &&\\
            \rhow(0)  &= \rho_{\text{wall}, 0},      &&\text{and} \\
            \vmJ{2}(t) &= 0                 &&t > 0.
        \end{aligned}
    \end{cases}
\end{equation}
Next, the time evolution of the effective strain and dislocation densities can be simulated using a time integrator. Because this often leads to a stiff system of equations, in \cref{sec:results}, we will use an implicit solver for numerically simulating the material evolution.

Following the state-space description from~\cite{furukawa1998}, we define a vector of inputs
\begin{equation}
    \bsx_{\text{raw}} \coloneqq (\varepsilon_\mathrm{vm}, \sigma_\mathrm{vm}, \T, \flux, \rhoc, \rhow)^T \in \bbR^6
\end{equation}
and a vector of outputs $\bsy_{\text{raw}} \coloneqq (\dot{\varepsilon}_\mathrm{vm}, \rhocdot, \rhowdot)^T \in \bbR^3$, such that the constitutive model can be written as $\bsy_{\text{raw}} = \calM(\bsx_{\text{raw}})$. For more efficient surrogate modeling, we apply a series of component-wise, invertible input and output transformations $\calT_{\text{in}}$ and $\calT_{\text{out}}$ to the raw model inputs and outputs, and reparametrize the model as
\begin{equation}
    \bsy_{\text{raw}} = \calT_{\text{out}}^{-1}(\calF(\calT_{\text{in}}(\bsx_{\text{raw}})).
\end{equation}
Here, $\calF(\;\cdot\;)$ represents a model in the transformed input and output space, i.e.,
\begin{equation}\label{eq:model}
    \bsy = \calF(\bsx)    
\end{equation}
with $\bsx \coloneqq \calT_{\text{in}}(\bsx_{\text{raw}}) \in \bbR^6$ and  $\bsy \coloneqq \calT_{\text{out}}(\bsy_{\text{raw}}) \in \bbR^3$. The exact forms of these transformations will be discussed in \cref{subsec:input_output_transformations}. In the remainder of this section, we will present two approaches for learning the model $\calF$ in \eqref{eq:model}.

\subsection{Response Surface Methodology}
\label{sec:response_surface_methodology}

In the Response Surface Methodology (RSM)~\cite{khuri2010}, we seek to find a functional relationship between the model output $\bsy = (y_i)_{i=1}^3$ and the input parameters $\bsx$ of the form
\begin{equation}\label{eq:response_surface_method}
	y_i(\bsx) = \bsPsi(\bsx)^T \bsalpha_i + \epsilon_i, \quad i = 1, \ldots, 3.
\end{equation}
Here, $\bsPsi(\,\cdot\,) : \bbR^6\rightarrow\bbR^p$ is a vector-valued function of the input parameters $\bsx$, $\bsalpha_i \in \bbR^p$ is a vector of unknown coefficients, and $\epsilon_j$ is a model error term. Often, $\bsPsi$ represents a mapping onto a low-order global polynomial basis, e.g., linear or quadratic. In this work, we assume $\bsPsi$ is a mapping of the input parameters onto a piecewise continuous linear function space.

Let $\Omega$ denote the input space, i.e., the domain over which we want to construct the constitutive surrogate, and decompose the domain into a mesh of non-overlapping elements $\Omega_e$ such that $\Omega = \bigcup_e \Omega_e$. The model output $y_i$ is approximated as
\begin{equation}
    y_i(\bsx) \approx \calF_{\text{RSM}, i}(\bsx) \coloneqq \sum_{j=1}^p \alpha_{i, j} \Psi_j(\bsx),
\end{equation}
where $\Psi_j$ is a basis function associated with the $j$th node of the mesh, and $\alpha_{i, j}$ is an unknown coefficient. We opt for locally-supported, linear continuous basis functions $\Psi_j$. In particular, the basis functions are constructed as the tensor product of one-dimensional functions
\begin{equation}
    \Psi_j(\bsx) = \prod_{k = 1}^6 \psi_{j, k}(x_k),
\end{equation}
with
\begin{equation}
    \psi_{j, k}(x_k) = 
    \begin{dcases}
        \frac{x_k - x_{j - 1, k}}{x_{j, k} - x_{j - 1, k}}, & \text{ if } x_{j - 1, k} < x_k < x_{j, k}, \\
        \frac{x_{j + 1, k} - x_k}{x_{j + 1, k} - x_{j, k}}, & \text{ if } x_{j, k} < x_k < x_{j + 1, k}, \\
        0, & \text{ otherwise},
    \end{dcases}
\end{equation}
where $x_{j, k}$ is the $k$th coordinate of the $j$th node, and $x_{j-1, k}$ and $x_{j + 1, k}$ are the coordinates of the nodes adjacent to node $j$ along the $k$th dimension.

The unknown coefficients $\bsalpha_i = (\alpha_{i, 1}, \ldots, \alpha_{i, p})^T$ can be found by minimizing the residual error across the entire domain, i.e., by solving
\begin{equation}
    \bsalpha_i^* = \argmin_{\bsalpha_i \in \bbR^p} \int_\Omega \bigg(y_i(\bsx) - \sum_{j=1}^p \alpha_{i, j} \Psi_j(\bsx)\bigg)^2 \dd\bsx,
\end{equation}
which can be converted into a least-squares problem by approximating the integral as the mean across the training points $\{(\bsx^{(n)}, y_i^{(n)})\}_{n=1}^N$, that is,
\begin{equation}\label{eq:rsm_least_squares}
    \bsalpha_{\text{LS}, i}^* = \argmin_{\bsalpha_i \in \bbR^p} \frac{1}{N} \sum_{n = 1}^N \bigg(y_i^{(n)} - \sum_{j=1}^p \alpha_{i, j} \Psi_j(\bsx^{(n)})\bigg)^2 = \argmin_{\bsalpha_i \in \bbR^p} \|\bsw - \Phi \bsalpha_i\|^2_2,
\end{equation}
where $\bsw = (y_i^{(1)}, \ldots, y_i^{(N)})^T$ is a vector with observed outputs, and $\Phi \in \bbR^{N \times p}$ is the design matrix, where each element $\Phi_{nj} = \Psi_j(\bsx^{(n)})$. Note that we also dropped the pre-factor $N^{-1}$ in the last equality, since it does not affect the minimizer. Since the design matrix $\Phi$ arising from piecewise polynomial basis functions defined over a mesh is inherently sparse, it is computationally efficient to use a direct solver for determining the unknown coefficients $\bsalpha_i$ from the normal equation
\begin{equation}\label{eq:normal_equations}
\bsalpha_i = (\Phi^T \Phi)^{-1} \Phi^T \bsw.
\end{equation}

\subsection{Mixture-of-experts surrogate models}
\label{sec:mixture_of_experts_surrogate_models}

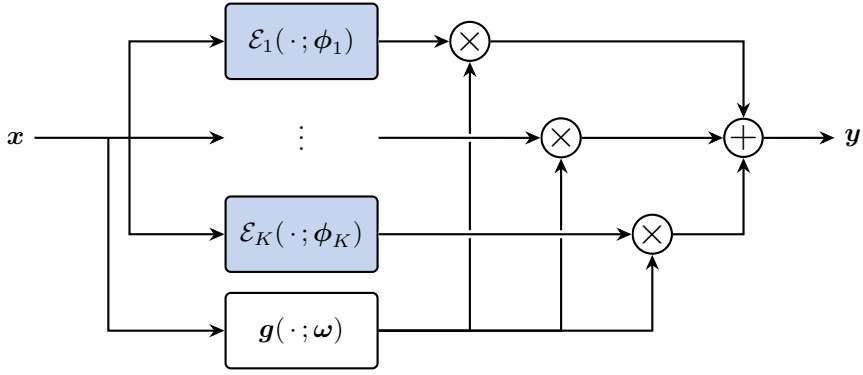
\begin{figure}
	\centering
	\input{fig/mixture_of_experts}
	\caption{Schematic overview of a mixture of experts architecture. The output is a weighted combination of the opinion of different experts $\calE_1(\bsx; \bsphi_1), \ldots, \calE_K(\bsx; \bsphi_K)$, with weights determined by the gating function $\bsg(\bsx; \bsomega)$.}
	\label{fig:mixture_of_experts}
\end{figure}

Motivated by the piecewise construction in \cref{sec:response_surface_methodology}, we now discuss a more general Mixture of Experts (MoE) architecture~\cite{jacobs1991, yuksel2012}. A MoE model consists of a collection of experts $\{\calE_k(\bsx; \bsphi_k)\}_{k=1}^K$ with parameters $\bsphi_k$, and a $K$-variate weighting function (also known as a gating function) $\bsg(\bsx; \bsomega) = (g_1(\bsx; \bsomega), \ldots, g_K(\bsx; \bsomega))^T$ with parameters $\bsomega$. The gating function combines the opinion of the different experts in an input-dependent fashion, to produce a single (multivariate) output
\begin{equation}\label{eq:mixture_of_experts}
	\bsy(\bsx) \approx \calF_\text{MoE}(\bsx) \coloneqq \sum_{k=1}^K g_k(\bsx; \bsomega) \calE_k(\bsx; \bsphi_k),
\end{equation}
see \cref{fig:mixture_of_experts}.

The gating functions are constrained such that $g_k(\bsx; \bsomega) \geq 0$ and
\begin{equation}
    \sum_{k=1}^K g_k(\bsx; \bsomega) = 1 \quad \text{for all } \bsx.
\end{equation}
These constraints ensure that the weights form a valid probability distribution over the $K$ experts, i.e., the weights can be interpreted as probabilities that a given expert $\calE_k$ will provide useful information for a given input, enabling the experts to specialize to particular inputs. Note also that, in a scalar output setting, the RSM from \cref{sec:response_surface_methodology} is a special case of~\eqref{eq:mixture_of_experts} where the (constant) experts are the coefficients $\bsalpha_i$, and the (deterministic and fixed) gating weights are the basis functions $\Psi_j$, controlling which experts contribute to the prediction at a given input $\bsx$.

Feed-forward neural networks have been proposed as good choices for the experts. A standard multilayer perceptron (MLP) neural network with $L$ layers can be written as
\begin{equation}\label{eq:feedforward_neural_network}
    \begin{aligned}
	\bsz_k^\ell &= \sigma(W_k^\ell \bsz_k^{\ell-1} + \bsb_k^\ell) \quad \text{ for } \ell = 1, 2, \ldots, L - 1 \\
	\bsz_k^L &= W_k^L z_k^{L-1} + \bsb_k^L,
    \end{aligned}
\end{equation}
where $\bsz_k^0$ is the input, $\bsz_k^L$ is the output, and $\sigma$ is a nonlinear activation function. The parameters of the MLP are the weight matrices and bias vectors, i.e., $\bsphi_k \coloneqq \{W_k^0, \bsb_k^0, \ldots, W_k^L, \bsb_k^L\}$. 

A convenient and commonly used expression for the gating function is the softmax function
\begin{equation}\label{eq:softmax}
	g_k(\bsx; \bsomega) = \bigg(\sum_{k=1}^K \exp(h_k(\bsx; \bsomega))\bigg)^{-1} \exp(h_k(\bsx; \bsomega)) \quad \text{ for } k = 1, 2, \ldots, K,
\end{equation}
where $h_k$ represents the gating value prior to the softmax operation. In a linear-softmax gating function, the $h_k$ are linear functions of the input $\bsx$, i.e.,
\begin{equation}\label{eq:linear_gating}
    h_k(\bsx; \bsomega) = \bsalpha_k^T\bsx + \beta_k,
\end{equation}
and where the weights $\bsomega \coloneqq \{\bsalpha_1, \beta_1, \ldots, \bsalpha_K, \beta_K\}$.

Given a dataset $\{(\bsx^{(n)}, \bsy^{(n)})\}_{n=1}^N$, the training objective for MoE models is to minimize a loss function defined over this data. For such a regression task, the most commonly used loss function is the mean squared error
\begin{equation}
    \calL(\bsphi, \bsomega) = \frac{1}{N} \sum_{n=1}^N \left( \bsy^{(n)} - \calF_\text{MoE}(\bsx^{(n)}) \right)^2.
\end{equation}
Remark that this results in a multivariate loss $\calL(\bsphi, \bsomega)$ across all outputs, which must be scalarized using a suitable reduction scheme (e.g., by taking the mean or sum). We use a mean reduction in \cref{sec:results}. Gradient-based optimization can subsequently be employed to minimize this loss with respect to both the expert parameters $\bsphi \coloneqq \{\bsphi_1, \ldots, \bsphi_K\}$ and the gating parameters $\bsomega$.

%% file: fig/creep_models.tex
\includegraphics{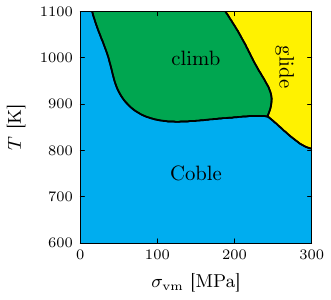}

%% file: fig/mixture_of_experts.tex
\newcommand{\sumjunc}{\Large$+$}
\newcommand{\prodjunc}{\Large$\times$}
\includegraphics{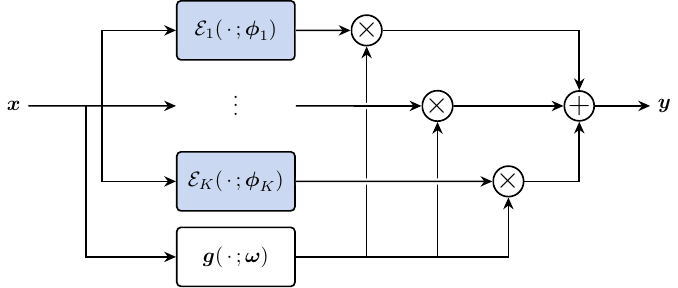}

%% file: src/results.tex
\section{Results}
\label{sec:results}

In this section, we present our main results obtained by applying the surrogate constitutive models developed in \cref{sec:methods} for simulating the viscoplastic creep behavior of \HT9{} steel. We start by discussing the generation of a simulation database that will serve as training data for our data-driven surrogate constitutive models. Next, we present the result of applying the RSM and MoE surrogates for our viscoplastic test problem, both in an offline setting and when deployed in a subsequent stress-strain prediction. We discuss and apply metrics for assessing the accuracy of these surrogates, and compare to a reference VPSC simulation.

\begin{table}
    \centering
    \begin{tabular}{c@{}lllll} \toprule
        & Parameter & Unit & Lower bound & Upper bound & Transform \\ \midrule
        & $\evm{}$ & - & 0 & 0.01 & $\log_{10}$, then scale to $(0, 1)$\\
        & $\vmJ2{}$ & \unit{\mega\pascal} & 0 & 300 & scale to $(-1, 1)$ \\ %
        \multirow{2}{*}{%
        } & $\T$ & \unit{\kelvin} & 600 & 1100 & scale to $(-1, 1)$ \\
        & $\flux{}$ & \unit{\dpa\per\second} & \num{1e-9} & \num{1e-6} & scale to $(-1, 1)$ \\ %
        \multirow{2}{*}{%
        } & $\rhoc{}$ & \unit{\per\square\metre} & \num{1e12} & \num{8.5e12} & scale to $(-1, 1)$ \\
        & $\rhow{}$ & \unit{\per\square\metre} & \num{5e12} & \num{12e12} & scale to $(-1, 1)$ \\ \bottomrule
    \end{tabular}
    \caption{Input parameters for the constitutive model with corresponding lower and upper bounds, as well as input transforms.}
    \label{tab:input_parameters}
\end{table}

\begin{table}
    \centering
    \begin{tabular}{c@{}lll} \toprule
         & Parameter & Unit & Transform \\ \midrule
         & $\evmdot{}$ & \unit{\per\second} & $\log_{10}$, then scale to $(0, 1)$ \\
         \multirow{2}{*}{%
         } & $\rhocdot{}$ & \unit{\per\square\metre\per\second} & Eq.~\eqref{eq:transformation} with $\kappa_{\text{cell}} = 10^{-10}, \eta = 0.3$ \\
         & $\rhowdot{}$ & \unit{\per\square\metre\per\second} & Eq.~\eqref{eq:transformation} with $\kappa_{\text{wall}} = 10^{-12}, \eta = 0.3$ \\ \bottomrule
    \end{tabular}
    \caption{Output parameters for the constitutive model with corresponding output transforms.}
    \label{tab:model_outputs}
\end{table}

\subsection{Database generation}
\label{subsec:database_generation}

To generate the training data required to fit our surrogate constitutive models, we run a series of VPSC simulations modeling the creep response of \HT{9}, see \Cref{sec:viscoplasticity_in_HT9_steel}. The \HT9{} steel texture, i.e., the crystallographic orientation distribution of grains within the material, is approximated by randomly assigning 50 uniformly distributed crystallographic orientations to ensure a representative distribution of grain orientations in the material. The random texture is kept the same across all simulations to isolate the effects of other variables being studied. This means our surrogate constitutive models are conditioned on this set of crystallographic orientations. The two active slip modes we consider in this work are
\begin{itemize}
    \item $\{110\}\langle111\rangle$: Slip along the $\langle111\rangle$ direction on the $\{110\}$ plane.
    \item $\{112\}\langle111\rangle$: Slip along the $\langle111\rangle$ direction on the $\{112\}$ plane.
\end{itemize}
These two modes account for a total of 24 slip systems (12 systems for each mode). The boundary conditions applied in the creep simulations are characterized by a prescribed, constant Cauchy stress tensor representative of a pressurized tube, and a constant temperature.

We sample the operating conditions and initial values for the strain and the dislocation densities according to a Latin Hypercube (LHS) design. This design ensures a broad and relatively homogeneous coverage of creep response of the material as a function of stress and temperature, thereby ensuring that the models can be trained to both capture regions dominated by one deformation mechanisms and transition regions in which more than one mechanism are activated to relatively similar levels. For each input sample, we run a full VPSC creep simulation, until a total accumulated effective plastic strain of 1\% has been reached, or until $10^9$ seconds have been simulated. For each simulation $k$, this results in a database of values ordered in an $N_k \times 9$ data matrix
\begin{equation}\nonumber
    D^{(k)} = \begin{pNiceArray}{ccccccIccc}
        \evm^{(k)}(t_0) & \vmJ2^{(k)}(t_0) & \T^{(k)} & \flux^{(k)} & \rhow^{(k)}(t_0) & \rhoc^{(k)}(t_0) & \evmdot^{(k)}(t_0) & \rhowdot^{(k)}(t_0) & \rhocdot^{(k)}(t_0) \\
        \evm^{(k)}(t_1) & \vmJ2^{(k)}(t_1) & \T^{(k)} & \flux^{(k)} & \rhow^{(k)}(t_1) & \rhoc^{(k)}(t_1) & \evmdot^{(k)}(t_1) & \rhowdot^{(k)}(t_1) & \rhocdot^{(k)}(t_1) \\
        \vdots & \vdots & \vdots & \vdots & \vdots & \vdots & \vdots & \vdots & \vdots \\
        \evm^{(k)}(t_{N_k}) & \vmJ2^{(k)}(t_{N_k}) & \T^{(k)} & \flux^{(k)} & \rhow^{(k)}(t_{N_k}) & \rhoc^{(k)}(t_{N_k}) & \evmdot^{(k)}(t_{N_k}) & \rhowdot^{(k)}(t_{N_k}) & \rhocdot^{(k)}(t_{N_k}) %
        \CodeAfter \UnderBrace[shorten, yshift=1.75mm]{last-1}{last-6}{\strut\text{inputs}} \UnderBrace[shorten, yshift=1.75mm]{last-7}{last-last}{\strut\text{outputs}}
    \end{pNiceArray},
    \vspace{2em}
\end{equation}
$k = 1, 2, \ldots, K_v$, where $N_k$ is the number of time steps in the $k$th simulation and $K_v$ is the total number of VPSC simulations. To prevent overfitting to specific output regimes and facilitate training, we create a more balanced data set by subsampling the rows of the data matrix $D^{(k)}$. In particular, we down-sample the time-dependent data by performing linear interpolation at 100 logarithmically spaced time steps between $t_0$ and $t_{N_k}$, see \cref{fig:subsampling} in \cref{sec:appendix} for more details. The motivation for this logarithmic subsampling is twofold. First, we use subsampling because the VPSC simulations capture a variety of physical behaviors, with some parameter combinations resulting in stiff systems that require very small time steps. Second, we use a logarithmic scaling because the dynamics of the system are mainly governed by the change in the dislocation densities, which is more important at small times. After logarithmic subsampling, all data matrices $D^{(k)}$ are concatenated to form a database with input and output data for training and testing. 
This results in about $9\times 10^5$ and $1.5\times 10^5$ number of training and testing samples, respectively.

\subsection{Input/output transformations}
\label{subsec:input_output_transformations}

As detailed in \cref{sec:viscoplasticity_in_HT9_steel}, we fit our surrogate model $\calF(\;\cdot\;)$ to predict the transformed outputs $\bsy = \calT_{\text{out}}(\bsy_{\text{raw}})$ given the transformed inputs $\bsx = \calT_{\text{in}}(\bsx_{\text{raw}})$. In this section, we detail the choice of input and output transform. 

The model inputs are transformed as follows. We apply a logarithmic transform in base 10 to the strain inputs $\evm$, and then linearly rescale to $(0, 1)$. We linearly rescale all other model inputs to $(-1, 1)$, see \cref{tab:input_parameters} and \cref{fig:input_transformations} in \cref{sec:appendix}. The model outputs are transformed as follows. We apply a logarithmic transform in base 10 to the effective strain rate outputs $\evmdot$, and then rescale to $(0, 1)$. The dislocation rate outputs $\rhocdot$ and $\rhowdot$ are transformed as
\begin{equation}\label{eq:transformation}
    \dot{\rho}_* \mapsto \sign(\dot{\rho}_*)|\kappa_* \dot{\rho}_*|^\eta \quad \text{for } * \in \{\text{cell}, \text{wall}\}
\end{equation}
with $\eta=0.3$, $\kappa_{\text{cell}}=10^{-10}$, and $\kappa_{\text{wall}}=10^{-12}$. Then, we shift the resulting values to the positive half-axis $(1, \infty)$, and apply a logarithmic transform in base 10, see \cref{tab:model_outputs} and \cref{fig:output_transformations} in \cref{sec:appendix}. These transformations are required because strain and dislocation rates depend in a nonlinear way on the operating conditions and model parameters, which prompts accuracy concerns when fitting the model response with the low-order polynomials that constitute the RSM method in \cref{sec:response_surface_methodology}. Further details are given in \cref{sec:appendix}.

\subsection{Comparison of Surrogate Constitutive Models}
\label{subsec:comparison_of_surrogate_constitutive_models}

We construct the RSM surrogate from \cref{sec:response_surface_methodology} for each of the three model outputs (the strain rate $\evmdot$ and the two dislocation density rates $\rhocdot$ and $\rhowdot$). To construct the response surface mesh, we discretize the input domain with 15~$\times$~15 connected elements in temperature and stress, and two elements in the strain dimension, along with single elements in the dimensions of dislocation density and irradiation dose rate, yielding a total of $p=768$ nodes. This discretization balances resolution and computational cost, and is guided by the expected variability of the deformation mechanisms across temperature and stress. The mesh is manually optimized during training to minimize deviations from the reference VPSC model while ensuring the surrogate does not overfit.

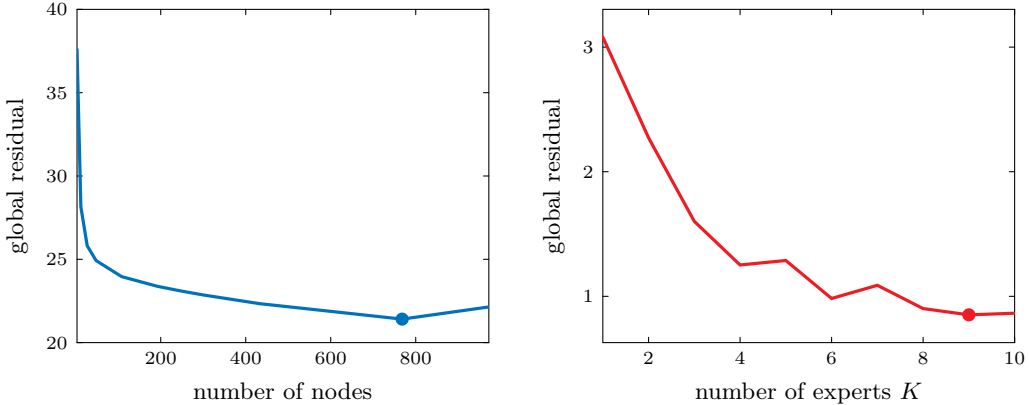
\begin{figure}
    \centering
    \begin{center}
        \input{fig/num_experts}
    \end{center}
    \caption{(\emph{left}) Global residual as a function of the number of nodes in the RSM method. The lowest residual of 21.4 is reached with 768 nodes total. (\emph{right}) Global residual as a function of the number of experts $K$, keeping all other (hyper)parameters constant. A residual of 0.85 is attained for $K=9$ experts.}
    \label{fig:num_expert}
\end{figure}

Next, we construct the MoE model from \cref{sec:mixture_of_experts_surrogate_models}. Each expert in the MoE surrogate is modeled as a dense neural network that is 4 layers deep and 64 neurons wide. Each layer is a concatenation of a linear layer, batch normalization and Gaussian Error Linear Units (GELU) activation functions. The use of GELU activation functions is motivated by their smoothness (i.e., they can be interpreted as continuous piecewise linear spline approximators), which may help improve convergence. The gating network, responsible for determining which expert to prioritize for a given input, is a linear network as defined in \cref{eq:softmax,eq:linear_gating}. We train the model for 10,000 epochs, using the Adam optimizer with decoupled weight decay regularization with weight decay coefficient $10^{-5}$~\cite{loshchilov2017}. The decoupled weight decay is employed to improve generalization. To dynamically adjust the learning rate and accelerate convergence, we use a cosine annealing scheduler with warm restarts~\cite{loshchilov2016}. The loss function minimizes the mean squared error between predictions and targets, with a mean reduction across all outputs. Hyperparameters such as number of layers, layer widths, activation functions and initial learning rates for expert and gating MLPs are determined using successive coordinate search~\cite{bergstra2012}.

One of the main challenges in MoE modeling is the determination of the number of experts $K$~\cite{yuksel2012}. We performed an exhaustive search over a discrete number of experts between $K=1$ and $K=10$, and found that 9 experts is a good trade-off between model complexity and test error minimization, see \cref{fig:num_expert}.

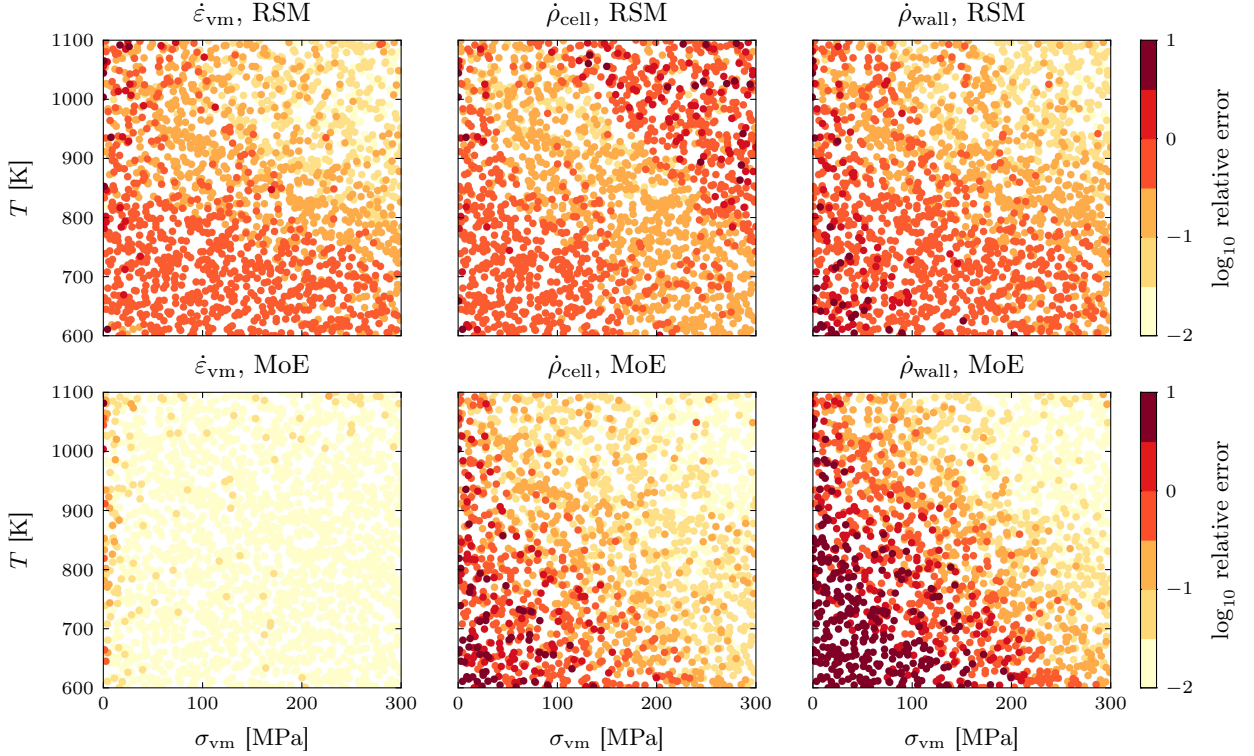
\begin{figure}[t]
    \begin{center}
        \input{fig/test_errors}
    \end{center}
    \caption{Relative test error for RSM surrogate constitutive model (\emph{top row}) and MoE surrogate constitutive model (\emph{bottom row}) as a function of $\vmJ2$ and $\T$ for all three constitutive model outputs (strain rate $\evmdot$ and dislocation density rates $\rhocdot$ and $\rhowdot$). Colors indicate the $\log_{10}$ relative error.}
    \label{fig:test_errors}
\end{figure}

\Cref{fig:test_errors} compares the ($\log_{10}$ of the) surrogate relative error on the test set for the RSM model (\emph{top row}) and the MoE model (\emph{bottom row}) as a function of $\vmJ2$ and $\T$ for all three constitutive model outputs. The reported errors are for the raw (back-transformed) outputs. Notice how the MoE model outperforms the RSM model in terms of accuracy for the strain rate $\evmdot$ across the entire ($\vmJ2{}, \T$)-space. The accuracy of the RSM model prediction for the strain rate appears to improve for large values of stress and temperature. For the dislocation density rates, the MoE model shows better accuracy than the RSM at higher stress and temperature values. However, the MoE accuracy deteriorates with decreasing stress and temperature, and in some cases it is less accurate than the RSM model.

\begin{figure}
    \centering
    \input{fig/error_metrics_sketch}
    \caption{Sketch of the error metrics used to assess the accuracy of the surrogate constitutive models: the relative error in the time-integrated strain (\emph{left}), the relative error in the final $\evm$ value (\emph{middle}), and the relative error in the rise time (\emph{right}).}
    \label{fig:error_metrics_sketch}
\end{figure}
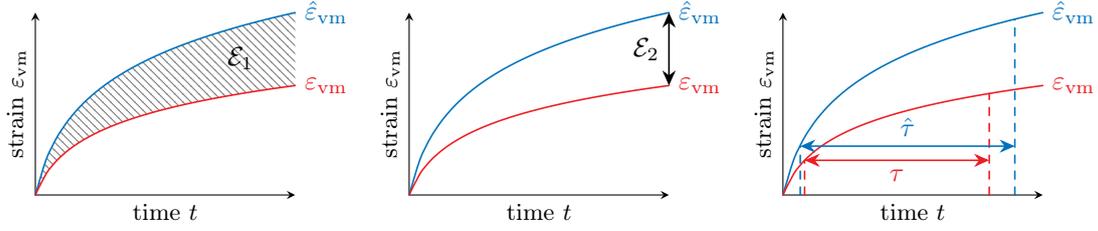

\begin{figure}
    \begin{center}
        \input{fig/error_metrics}
    \end{center}
    \caption{Relative error for RSM surrogate constitutive model (\emph{left column}) and MoE surrogate constitutive model (\emph{middle column}) as a function of $\vmJ2$ and $\T$, for time-integrated strain (\emph{top row}), the final strain value (\emph{middle row}), and the rise time. Color scale indicates the $\log_{10}$ relative error. The \emph{right column} indicates the difference between the two errors. Negative values (red) indicate where the RSM surrogate constitutive model performs better than the mixture of experts model.}
    \label{fig:error_metrics}
\end{figure}
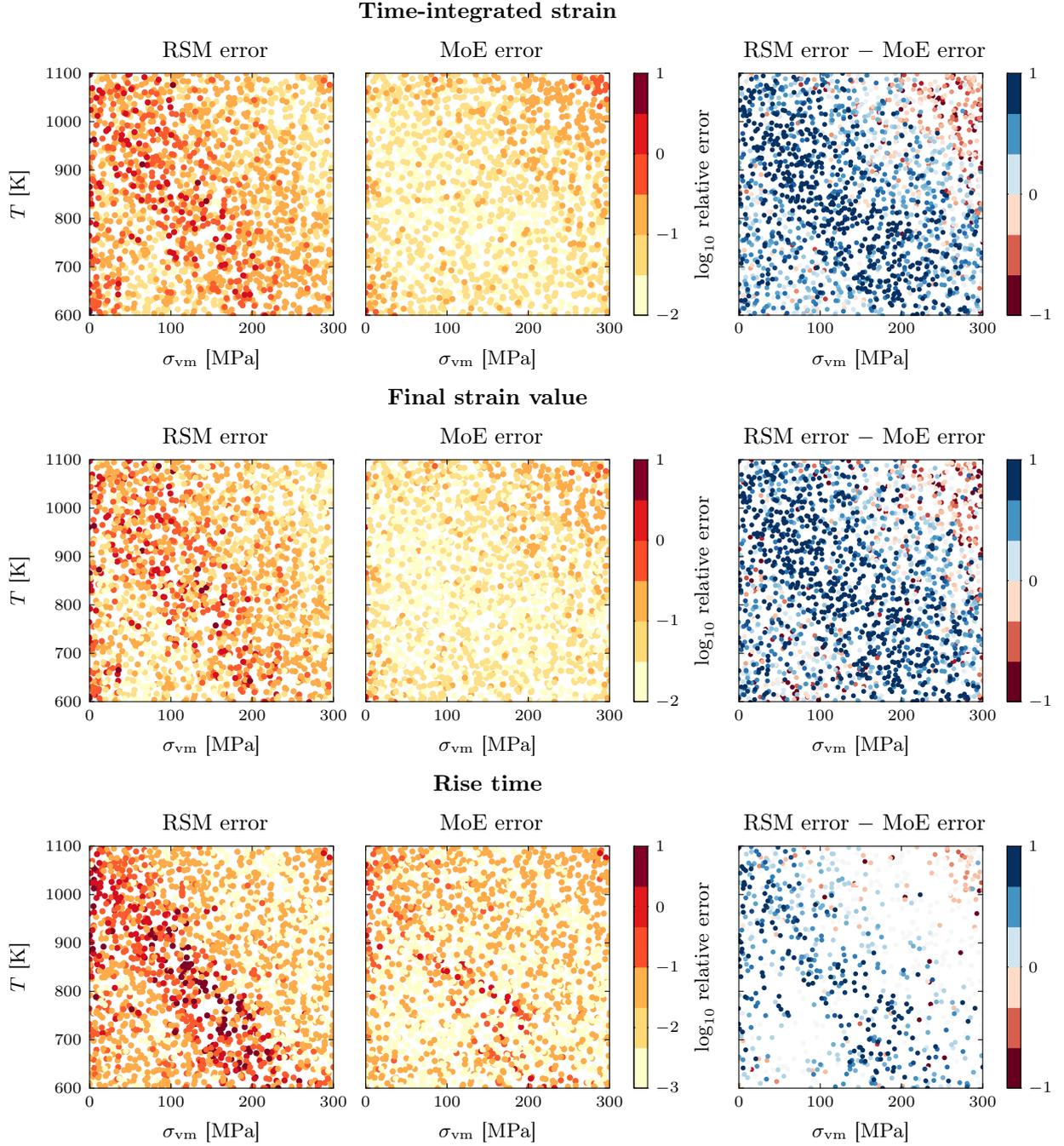

Next, we investigate how both surrogate constitutive models perform in an online setting, i.e., when they are used to replace the right-hand side of \eqref{eq:normality_rule} to simulate the creep response of \HT{9}. To numerically solve the system, we employ the \texttt{LSODA} solver from SciPy, which is a wrapper for the Fortran solver from ODEPACK~\cite{hindmarsh1983}. For the RSM constitutive model, an additional complexity is that one cannot evaluate the model for parameter combinations outside the range specified by the input parameters, see~\cref{tab:input_parameters}. In such cases, we effectively clipped the model inputs to remain strictly inside the specified bounds. The MoE model did not impose such a constraint, and we evaluated the constitutive model as is. We remark that the MoE model appears to predict physically consistent results, although we do not have validation data in this regime. Solving \eqref{eq:normality_rule} results in a time-dependent approximation $\hat\varepsilon_\text{vm}(t)$ for the strain $\evm(t)$, which we can compare against the strain predicted by the VPSC simulations. To quantify the error in the approximation, we define the following error metrics (see \cref{fig:error_metrics_sketch}):
\begin{enumerate}
    \item The relative error in the time-integrated strain, defined as
    \begin{equation}
        \calE_1 \coloneqq \frac{I(\hat\varepsilon_\text{vm}(t) - \evm(t))}{I(\evm(t))} \quad \text{with} \quad I(f) = \int_0^{t_\text{end}} f^2(t) \mathrm{d}t
    \end{equation}
    where $t_\text{end}$ is the last simulated time.
    \item The relative error in the final $\evm$ value
    \begin{equation}
        \calE_2 \coloneqq \frac{|\hat\varepsilon_\text{vm}(t_\text{end}) - \evm(t_\text{end})|}{|\evm(t_\text{end})|}.
    \end{equation}
    \item The relative error in the rise time, i.e., the duration it takes for the strain to transition from 10\% to 90\% of its steady-state value, defined as
    \begin{equation}
        \calE_3 \coloneqq \frac{|\hat{\tau} - \tau|}{|\tau|}
    \end{equation}
    with $\tau = t_{90} - t_{10}$, and where the time $t_k$ is such that
    \begin{equation}
        \evm(t_k) = \evm^k  \quad  \text{and} \quad \evm(t) < \evm^k \text{ for } t < t_k,
    \end{equation}
    with $\evm^k = k \% \cdot \evm(t_\text{end})$.
\end{enumerate}
We evaluate these error metrics for both the RSM and MoE models. The results are shown in \cref{fig:error_metrics}. For the majority of error metrics and stress and temperature values, the MoE constitutive model outperforms the RSM constitutive model in terms of accuracy of the predicted strain $\hat\varepsilon_\text{vm}(t)$. The MoE strain predictions are slightly less accurate for large values of $\vmJ{2}$ and $\T$.

\begin{figure}
    \centering
    \input{fig/stress_strain}
    \caption{A comparison between the RSM (\raisebox{1.5pt}{\protect\tikzexternaldisable{}\protect\tikz{\protect\draw[default line, very thick, line cap=round, Color1] (0, 0) -- (0.33, 0);}}) and MoE (\raisebox{1.5pt}{\protect\tikzexternaldisable{}\protect\tikz{\protect\draw[default line, very thick, line cap=round, Color2] (0, 0) -- (0.33, 0);}}) surrogate constitutive models and the validation data when predicting the strain $\evm$ for a dose rate $\flux = \SI[per-mode=symbol]{1e-8}{\dpa\per\second}$. We also indicate the different creep mechanisms from \cref{fig:creep_models}: glide (\raisebox{-0.25pt}{\protect\tikzexternaldisable{}\protect\tikz{\protect\fill[Yellow] (0, 0) rectangle (0.2, 0.2);}}), climb (\raisebox{-0.25pt}{\protect\tikzexternaldisable{}\protect\tikz{\protect\fill[ForestGreen] (0, 0) rectangle (0.2, 0.2);}}) and Coble (\raisebox{-0.25pt}{\protect\tikzexternaldisable{}\protect\tikz{\protect\fill[Cyan] (0, 0) rectangle (0.2, 0.2);}}). A {\protect\tikzexternaldisable{}\protect\tikz{\protect\node[draw=Color1, fill=white, very thick, star, star points=5, star point ratio=1.7, scale=0.4] {};}} indicates the last value that can be predicted with the RSM surrogate, which is restricted to the support of the training data.}
    \label{fig:stress_strain}
\end{figure}

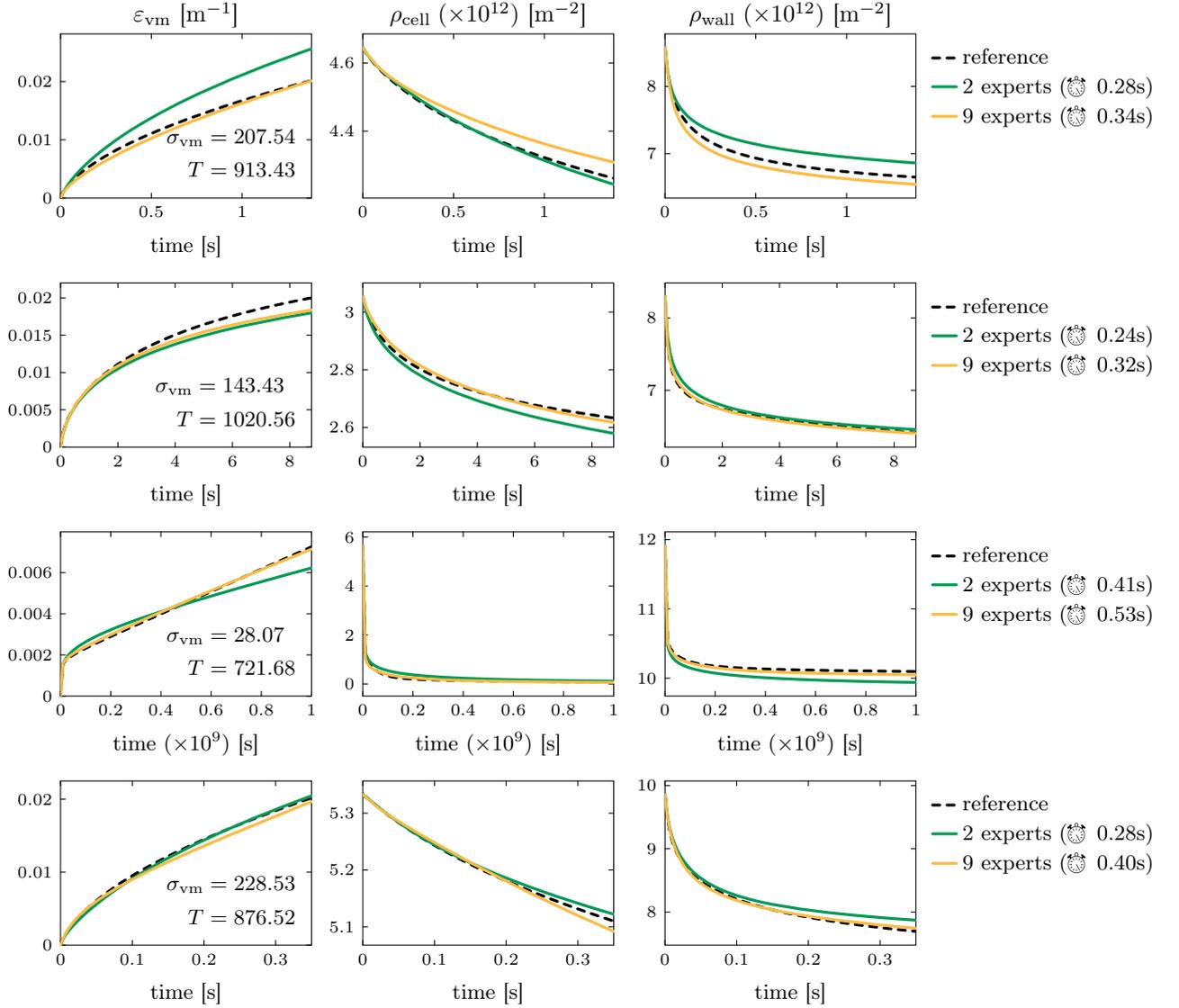
\begin{figure}
    \centering
    \input{fig/moe_accuracy}
    \caption{Time-integrated strain and material parameters for the data-driven constitutive MoE model with 9 (\raisebox{1.5pt}{\protect\tikzexternaldisable{}\protect\tikz{\protect\draw[default line, very thick, line cap=round, Color4] (0, 0) -- (0.33, 0);}}) and 2 (\raisebox{1.5pt}{\protect\tikzexternaldisable{}\protect\tikz{\protect\draw[default line, very thick, line cap=round, Color3] (0, 0) -- (0.33, 0);}}) experts. The dashed line indicates the reference from the validation data. Clocks ({\small\raisebox{-1.5pt}{\scriptsize\StopWatchEnd}}{}) indicate the best wall clock time over 1\,000 repetitions of the forward simulation.}
    \label{fig:moe_accuracy}
\end{figure}

%% file: fig/num_experts.tex
\includegraphics{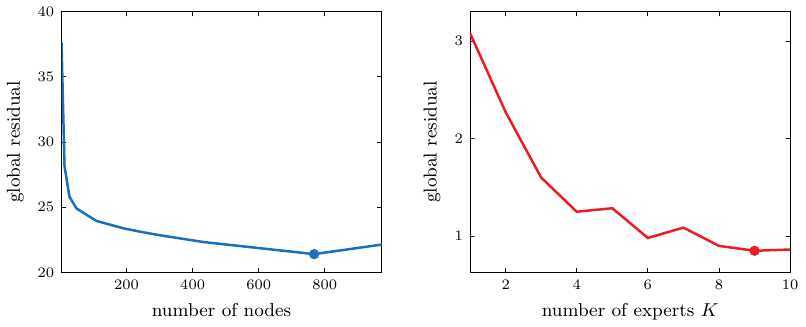}

%% file: fig/test_errors.tex
\includegraphics{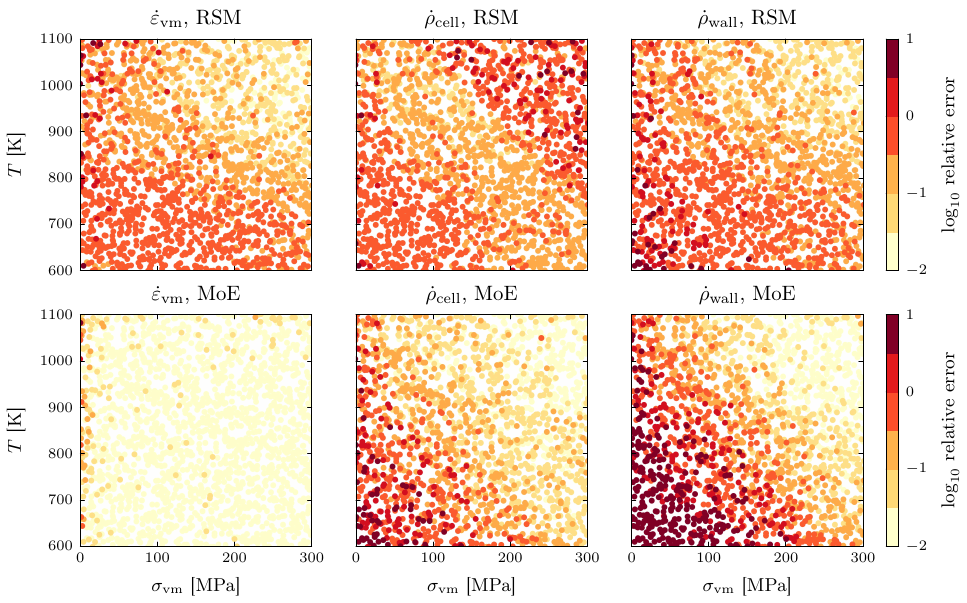}

%% file: fig/error_metrics_sketch.tex
\includegraphics{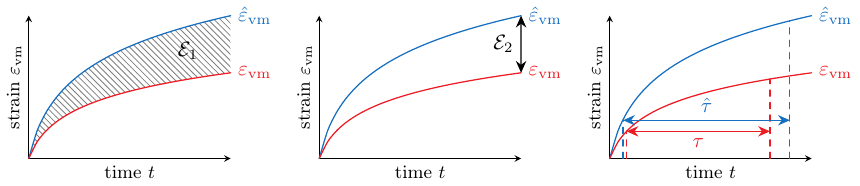}

%% file: fig/error_metrics.tex
\includegraphics{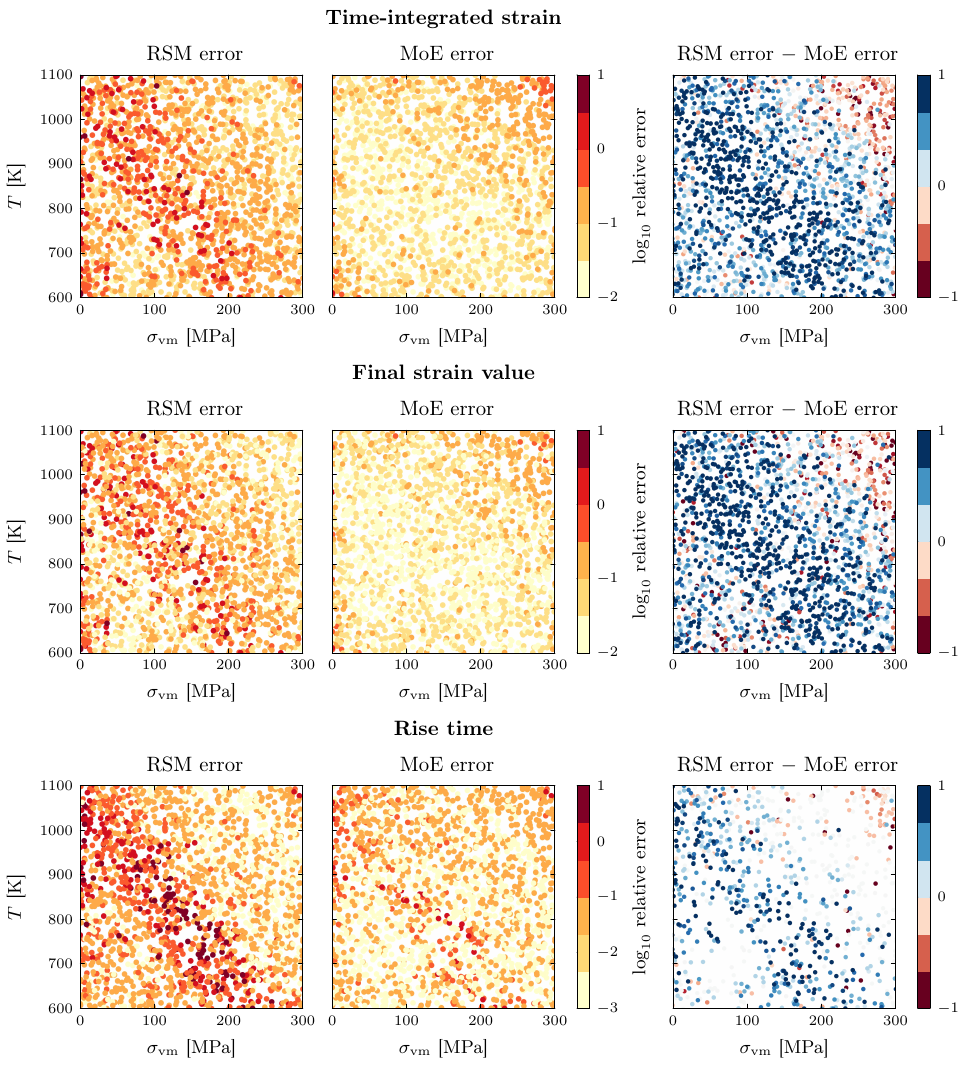}

%% file: fig/stress_strain.tex
\includegraphics{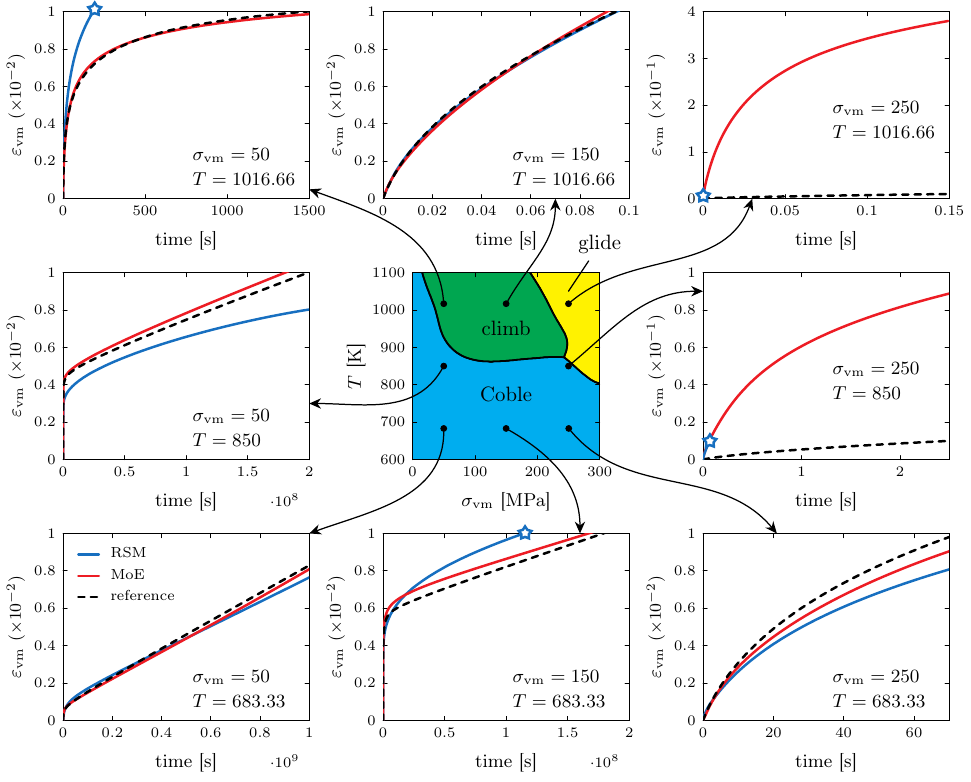}

%% file: fig/moe_accuracy.tex
\includegraphics{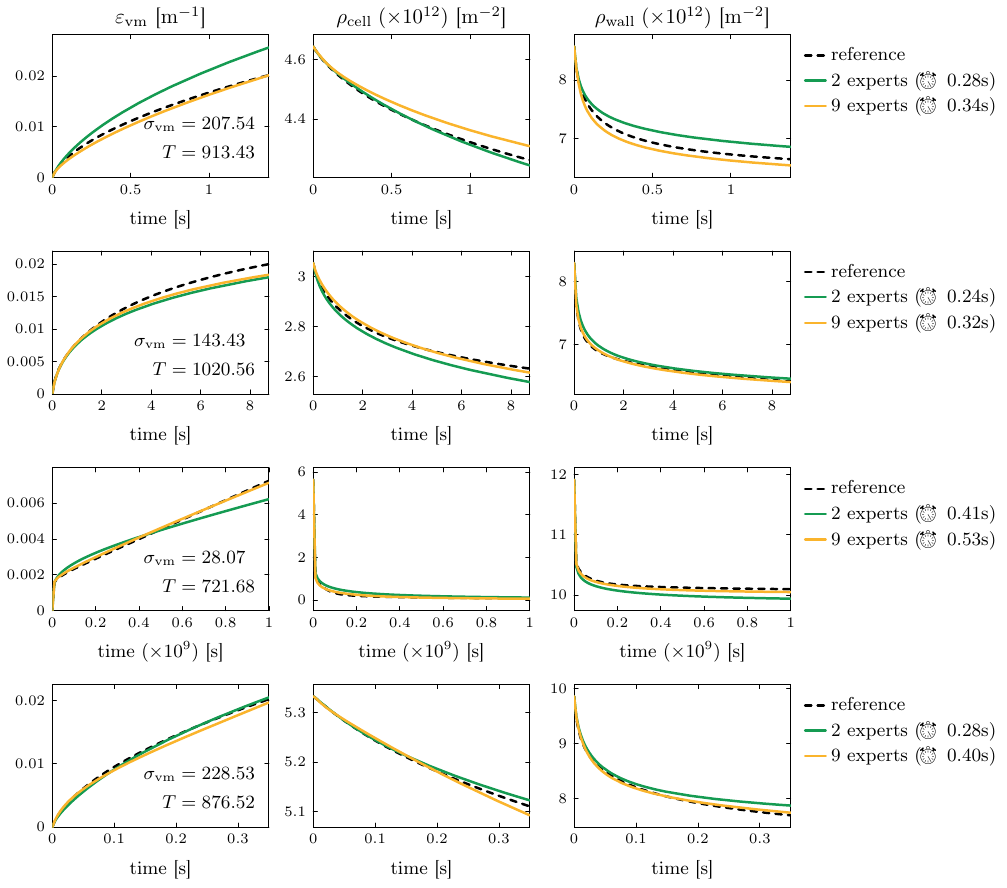}

%% file: src/discussion.tex
\section{Discussion}
\label{sec:discussion}

We simulate the time evolution of the creep response of the polycrystalline sample, with a constant stress imposed on the system. \Cref{fig:stress_strain} shows the strain predictions as a function of time for the RSM and MoE models for selected values of $\vmJ2{}$ and $\T$, and compares them to the reference VPSC predictions, assuming an irradiation dose rate of $\flux = 10^{-8} \unit{\dpa\per\second}$. This figure again confirms that the MoE predictions of the strain are, in general, more accurate than the corresponding RSM predictions. Furthermore, the predictions obtained from the RSM surrogate are restricted to the interpolation regime (effective strain values below $10^{-2}$). This is indicated in \cref{fig:stress_strain} by the {\tikzexternaldisable{}\tikz{\node[draw=Color1, fill=white, very thick, star, star points=5, star point ratio=1.7, scale=0.4] {};}} marker.

We observe a good agreement between the MoE prediction and the reference simulations across the $\T$ and $\vmJ{2}$ space, except for large values of temperature and stress (corresponding to the glide regime). This is consistent with the results shown in \cref{fig:test_errors}. Notably, the RSM and MoE constitutive models predict similar values at early times (effective strain values below $10^{-2}$), indicating an issue with the training data rather than the surrogate construction approach. The glide regime is governed by complex, mechanistically rich behavior arising from the dynamic interplay of dislocation generation, annihilation and trapping, and is characterized by rapidly evolving microstructures and high strain rates. Consequently, the underlying high-fidelity models used to train the surrogates are inherently less accurate in this regime.

We remark that the errors in the predicted strain values are due to the accumulation of surrogate error during the solution of \eqref{eq:model}. To further clarify this point, \cref{fig:moe_accuracy} shows the strain predictions for the MoE model with two different models: the original model with 9 experts, and a model with $K = 2$ experts (see \cref{fig:num_expert} for the corresponding increase in the global residual). As shown in \cref{fig:moe_accuracy}, the low-fidelity model with 2 experts accumulates more error in the strain $\evm$ over time, but predictions remain acceptable in terms of accuracy. Also shown in \cref{fig:moe_accuracy} is the effect of the reduction in the number of experts on the wall clock time for simulating the material behavior across time. Restricting the model from 9 to 2 experts reduces the wall clock time by 20 to 30\%, allowing us to trade accuracy for computational cost accordingly. For comparison, the corresponding VPSC simulations for these reference scenarios took between 15 minutes and 2 hours, while the surrogates complete simulations within a fraction of a second.

\begin{figure}
    \centering
    \input{fig/irradiation}
    \caption{A comparison between the RSM (\raisebox{1.5pt}{\protect\tikzexternaldisable{}\protect\tikz{\protect\draw[default line, very thick, line cap=round, Color1] (0, 0) -- (0.33, 0);}}) and MoE (\raisebox{1.5pt}{\protect\tikzexternaldisable{}\protect\tikz{\protect\draw[default line, very thick, line cap=round, Color2] (0, 0) -- (0.33, 0);}}) surrogate constitutive models and the validation data when predicting the strain $\evm$ for an irradiation dose rate $\flux = \SI[per-mode=symbol]{5e-7}{\dpa\per\second}$. This figure is the equivalent of the 3 panes in the top left corner of \cref{fig:stress_strain}, but at a higher irradiation dose rate value. Note how the performance of the surrogate models degrades for larger dose rate values. A {\protect\tikzexternaldisable{}\protect\tikz{\protect\node[draw=Color1, fill=white, very thick, star, star points=5, star point ratio=1.7, scale=0.4] {};}} indicates the last value that can be predicted with the RSM surrogate, which is restricted to the support of the training data.}
    \label{fig:irradiation}
\end{figure}

To investigate the effect of irradiation (i.e., the irradiation dose rate $\flux$), we repeat the strain predictions from \cref{fig:stress_strain} in \cref{fig:irradiation}, with a larger irradiation dose rate $\flux = \SI[per-mode=symbol]{5e-7}{\dpa\per\second}$. As in this model, irradiation only has an effect when the climb mechanism is dominant, only 3 panes from \cref{fig:stress_strain} are repeated. We note that both the RSM and MoE surrogate models struggle to replicate the reference VPSC simulations for this larger dose rate. Note again how the prediction of the RSM surrogate constitutive model are restricted to effective strain values below $10^{-2}$, as indicated by the {\tikzexternaldisable{}\tikz{\node[draw=Color1, fill=white, very thick, star, star points=5, star point ratio=1.7, scale=0.4] {};}} markers. 

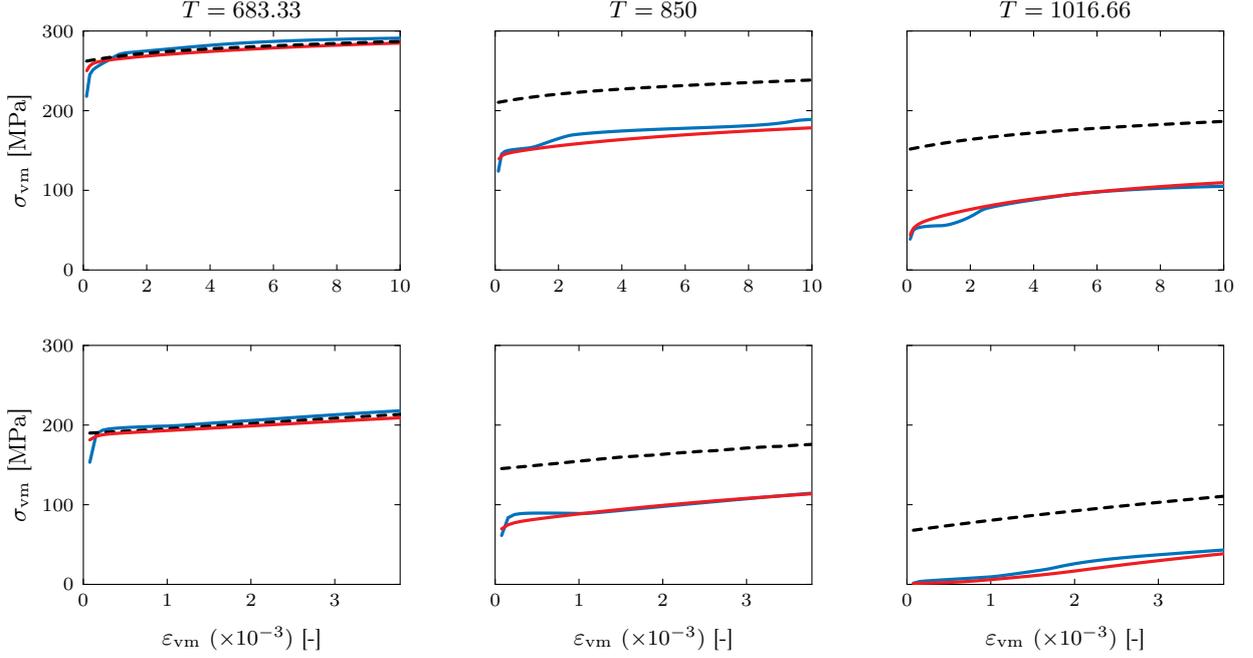
\begin{figure}
    \centering
    \input{fig/tensile}
    \caption{Comparison between the RSM (\raisebox{1.5pt}{\protect\tikzexternaldisable{}\protect\tikz{\protect\draw[default line, very thick, line cap=round, Color1] (0, 0) -- (0.33, 0);}}) and MoE (\raisebox{1.5pt}{\protect\tikzexternaldisable{}\protect\tikz{\protect\draw[default line, very thick, line cap=round, Color2] (0, 0) -- (0.33, 0);}}) surrogate constitutive models and the validation data for a tensile test at different temperatures and an imposed strain rate of \SI{1e-3}{\per\second} (\textit{top}) and \SI{1e-5}{\per\second} (\textit{bottom}).}
    \label{fig:tensile}
\end{figure}

Next, we compare the performance of our surrogate constitutive models (trained on data from a creep loading case only) for a tensile loading case. Under tensile loading, the material is subjected to a uniaxial deformation driven by Dirichlet boundary conditions imposing a constant axial strain rate. In contrast, the previously described creep load case applies a constant axial stress. The tensile load case induces an immediate elastic response, followed by viscoplastic deformation, resulting in a total strain that increases linearly with time under the imposed constant strain rate. The results are shown in \cref{fig:tensile}, for an imposed strain rate of $\dot{\varepsilon} = \SI{1e-3}{\per\second}$ (\textit{top}) and $\dot{\varepsilon} = \SI{1e-5}{\per\second}$ (\textit{bottom}). Note that the MoE and RSM surrogate models are able to capture the trend (shape) in the stress-strain curves, but a temperature-dependent bias is present that is independent of the imposed strain rate, except at low temperatures. These low temperatures correspond to regions in \cref{fig:test_errors} and \cref{fig:error_metrics} where we obtained good accuracy. However, this is no longer the case at higher temperatures where the model accuracy typically decreases. This indicates that, should we want to capture both the creep and tensile loading behavior of the material, we would need to enrich our database with data from VPSC simulations in a tensile regime.

\begin{figure}
    \begin{center}
        \input{fig/expert_domains}
    \end{center}
    \caption{Illustration of the different domains generated by the MoE model when evaluating the test data. Different experts are represented by different colors. Shading indicates the degree of trust in the individual experts. Dashed lines indicate regions where some of the lesser-used experts are active. Note that a third parameter, the dose rate $\flux$, is not shown on this picture.}
    \label{fig:expert_domains}
\end{figure}
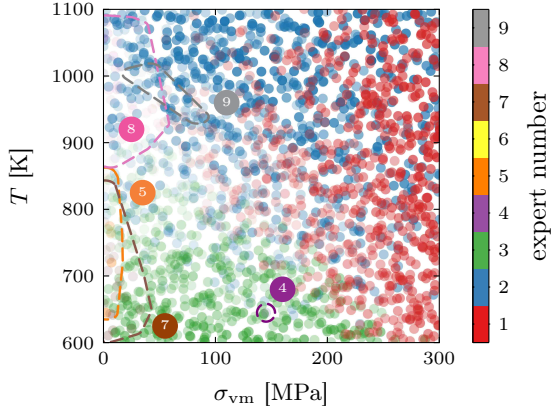

One of the advantages of the MoE approach over more traditional neural architectures is that a certain degree of locality can be recovered. In particular, for a given temperature and stress value, we can identify which of the expert's opinions is trusted the most, identifying the localized behavior inherently observed by the MoE model. To this end, we inspect the output values of the gating function $\bsg$ in \eqref{eq:mixture_of_experts} as we evaluate the model on the test data. The expert with the most valued opinion will be associated with the largest gating function output value. The results are shown in \cref{fig:expert_domains}, where the opinion of different experts is represented by different colors. The shading indicates the degree of trust in the opinion of that expert alone (i.e., the value of the gating function). Dashed lines indicate the region where some of the lesser-used experts are active. Note how most of the $(\vmJ2{}, \T)$-plane can be covered by the output of only a handful, e.g., 3 to 4, experts. Other experts, such as experts number 4 and 9, are only responsible for a small, localized region of the input parameter space. This may in part explain the behavior of the residual of the MoE model as a function of the number of experts shown in \cref{fig:num_expert}, where we observe a plateau in residual (loss) value for $K > 3$ experts. Also note that, in these plots, there is a hidden third parameter, the irradiation dose rate $\flux$, that is not shown in the picture, which in part explains the overlap between the experts.

Comparing \cref{fig:creep_models,fig:expert_domains} we observe that the different domains generated by the MoE model to some degree coincide with the dominant creep mechanisms identified in \cref{sec:viscoplasticity_in_HT9_steel}. Finally, we remark that the MoE framework automatically assigns experts to certain regions in parameter space, as opposed to the tiling approach of the RSM constitutive model.

%% file: fig/irradiation.tex
\includegraphics{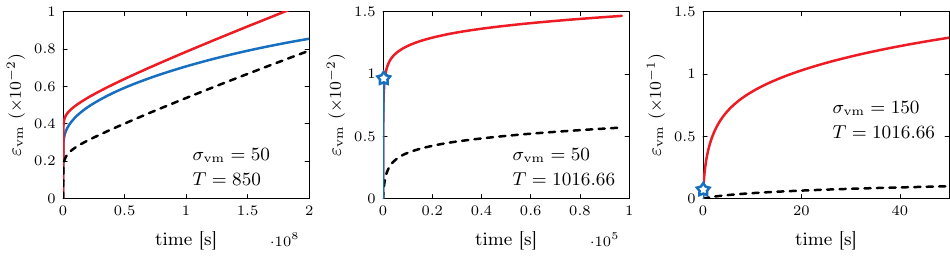}

%% file: fig/tensile.tex
\includegraphics{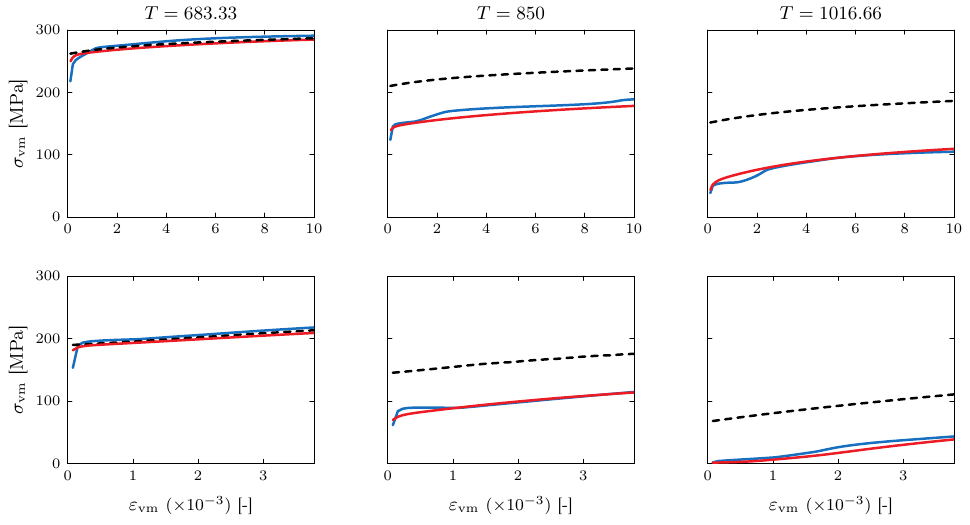}

%% file: fig/expert_domains.tex
\includegraphics{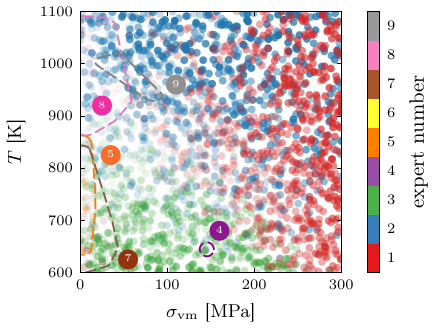}

%% file: src/conclusion.tex
\section{Conclusion}
\label{sec:conclusion}

Motivated by recent successes in the mechanics of materials community, where constitutive models that use internal state variables (dislocation densities) as a metric for the state of the microstructure can be used to predict complex phenomena such as strain hardening as well as softening response, we developed two novel local surrogate constitutive models for viscoplasticity in \HT{9} steel. The proposed methods adopt a localized modeling strategy, in which the input space is partitioned and separate surrogate models are trained for different regions. These regions are defined either explicitly in the case of the piecewise polynomial construction in the response surface method (RSM), or implicitly through the gating network in the mixture of experts (MoE) model. While in the current application, the two frameworks use different local models, in general, the RSM approach can be thought as a special case of MoE for which the gating functions are predefined and fixed.

We trained our surrogate models on high-fidelity VPSC data generated under creep loading and compared their ability to predict strain rate and dislocation density rates across a broad range of temperatures, stresses, and irradiation dose rates. Our results show that the MoE model consistently outperforms the RSM surrogate in predictive accuracy. The predictions then serve as input to a forward simulation in which the constitutive model is embedded in an ODE solver to evolve the strain over time. To assess accuracy, we compared the strain predictions obtained from these surrogate-driven simulations to those from full VPSC simulations. We introduced three physics-informed error metrics (relative error in strain rate, relative error in final strain, and relative error in rise time) to quantify discrepancies in the resulting strain curves. These metrics capture both instantaneous and time-integrated response errors. We found that while both surrogates perform well under moderate loading conditions, prediction errors increase significantly at high temperatures and stresses. Detailed comparisons in the climb, Coble, and glide regimes reveal that both models struggle in the dynamically complex glide regime, whereas the MoE model maintains good accuracy in the climb and Coble regimes.

Given that full-field VPSC simulations can take between 15 minutes and 2 hours, the computational speed-up provided by our surrogate models, producing results in under a second, is substantial. However, surrogate accuracy was observed to degrade at higher dose rates in the climb regime. This is likely due to the limited representation of high-dose microstructures in the training data, which leads to poor generalization in regions where irradiation-induced hardening plays a dominant role.

Finally, in attempt to understand if constitutive model surrogates tracking internal state variables can be used to extrapolate to unseen loading scenarios, we tested both surrogate models under tensile loading. While typical surrogates are expected to interpolate only, the use of these internal state variables motivates testing the constitutive model for extrapolation. We found that the model will in all cases tested perform extremely well in predicting the strain hardening of the material (under creep loading), however, the surrogates do not perform well when dealing with the yield strength (under tensile loading), except at low temperatures. The resulting stress-strain curves showed consistent bias in both RSM and MoE predictions, suggesting that including tensile data during training is essential to achieve reliable extrapolation beyond the creep loading conditions used for calibration.

In future work, we will expand the training dataset to include a more diverse and balanced set of loading paths, including tensile, cyclic, and multi-axial loading, and extend the MoE architecture with physics-informed regularization to improve interpretability and robustness in data-sparse regions. Furthermore, integration of uncertainty quantification techniques will be explored to support the use of these surrogates in probabilistic simulations and design applications.

%% file: src/acknowledgements.tex
\section*{Acknowledgements}

This work was supported by the U.S. Department of Energy, Office of Nuclear Energy, and Office of Science, Office of Advanced Scientific Computing Research through the Scientific Discovery through Advanced Computing project on Simulation of the Response of Structural Metals in Molten Salt Environment.

This research made use of Idaho National Laboratory's High Performance Computing systems located at the Collaborative Computing Center and supported by the Office of Nuclear Energy of the U.S. Department of Energy and the Nuclear Science User Facilities under Contract No.\,DE-AC07-05ID14517.

This article has been co-authored by employees of National Technology and Engineering Solutions of Sandia, LLC under Contract No.\,DE-NA0003525 with the U.S. Department of Energy (DOE). The employees co-own right, title and interest in and to the article and are responsible for its contents. The United States Government retains and the publisher, by accepting the article for publication, acknowledges that the United States Government retains a non-exclusive, paid-up, irrevocable, world-wide license to publish or reproduce the published form of this article or allow others to do so, for United States Government purposes. The DOE will provide public access to these results of federally sponsored research in accordance with the DOE Public Access Plan \url{https://www.energy.gov/downloads/doe-public-access-plan}.

Los Alamos National Laboratory, United States, an affirmative action/equal opportunity employer, is operated by Triad National Security, LLC, for the National Nuclear Security Administration of the U.S. Department of Energy under Contract No. 89233218CNA000001.

%% file: src/appendix.tex
\section{Input/output transformations}\label{sec:appendix}

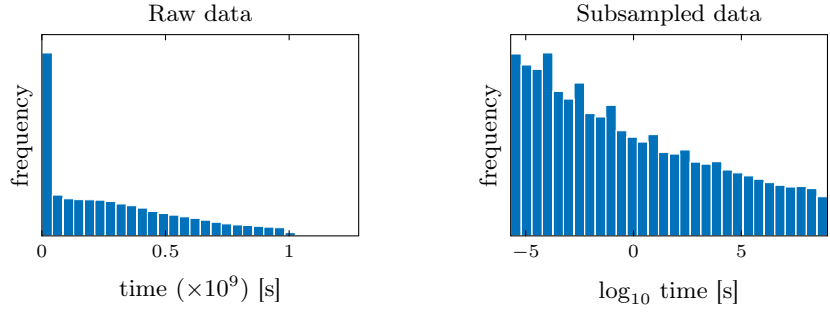
\begin{figure}
    \centering
    \input{fig/downsampling}
    \caption{Effect of subsampling on the selected time-indices for the training data.}
    \label{fig:subsampling}
\end{figure}

\begin{figure}
    \centering
    \input{fig/output_histograms}
    \caption{Effect of subsampling and transformations for the output data, see \cref{tab:model_outputs}. The top row shows the distribution of the raw output data, and the bottom row shows the distribution of the transformed output data.}
    \label{fig:output_transformations}
\end{figure}
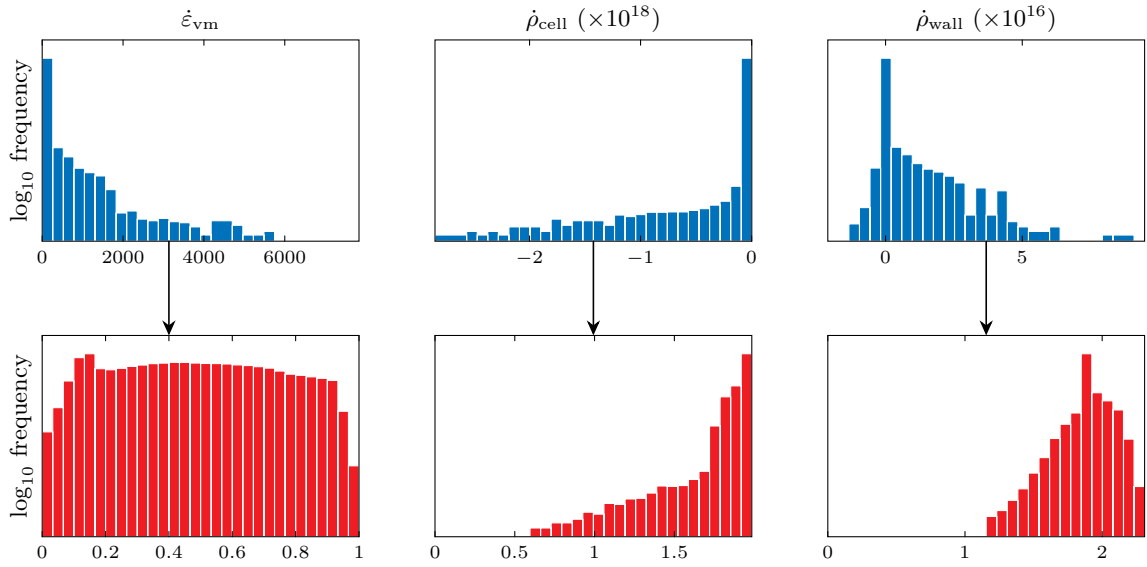

\Cref{fig:subsampling} shows the effect of the subsampling of the raw simulation data on the distribution of the retained time instances. \Cref{fig:input_transformations} shows the effect of subsampling and input transformations on the retained input data. After transformation, the input training data is much more evenly distributed across the input range. \Cref{fig:output_transformations} shows the effect of subsampling and output transformations on the retained output data. Note that the top row with raw output samples shows the $\log_{10}$ of the frequency. We found the output transformations to be a critical addition to achieve high accuracy of the surrogate models. 

\begin{figure}
    \centering
    \input{fig/input_histograms}
    \caption{Effect of subsampling and transformations for the input data, see \cref{tab:input_parameters}. Rows 1 and 3 show the distribution of the raw input data, and rows 2 and 4 show the distribution of the transformed input data.}
    \label{fig:input_transformations}
\end{figure}
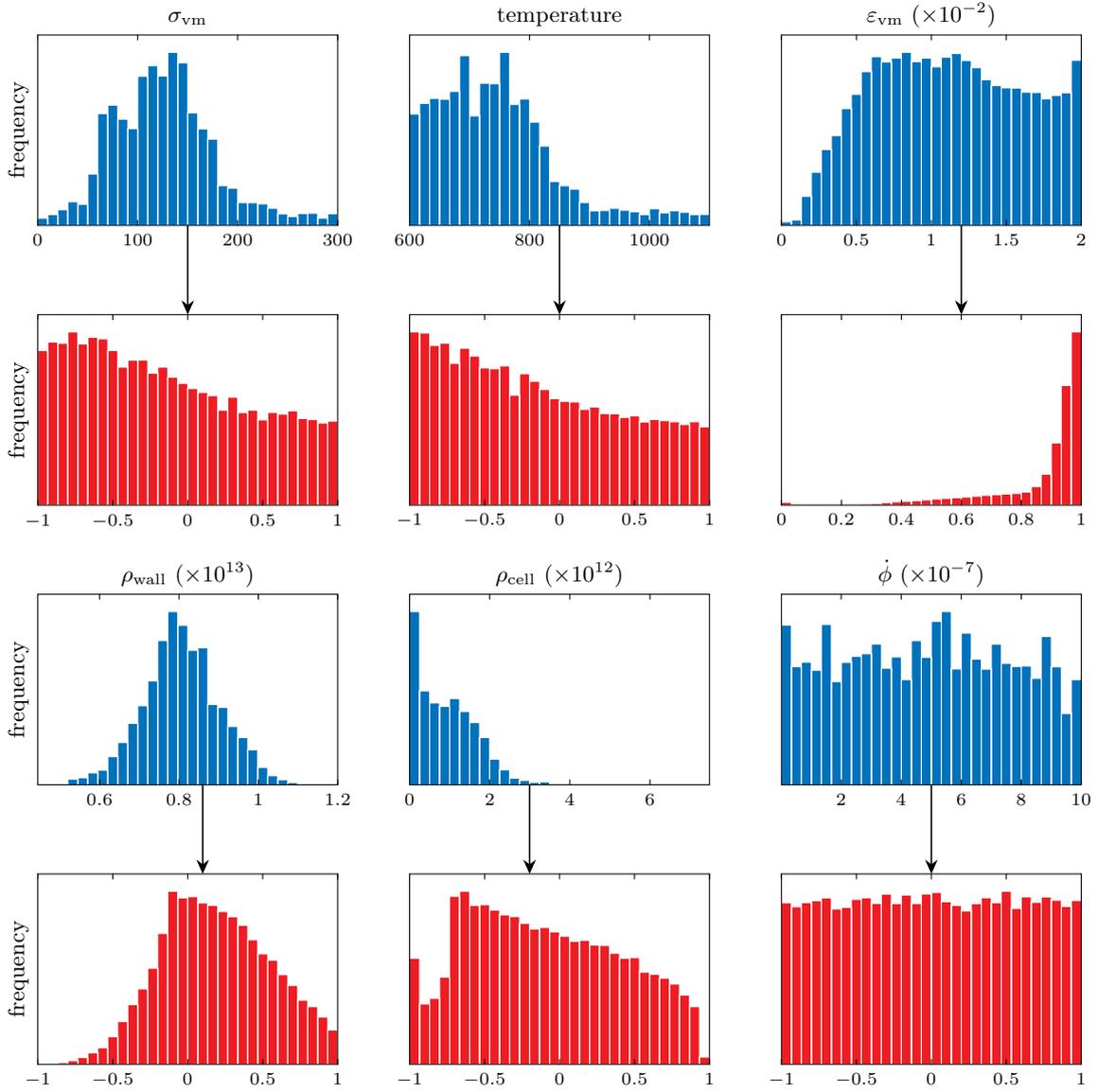

%% file: fig/downsampling.tex
\includegraphics{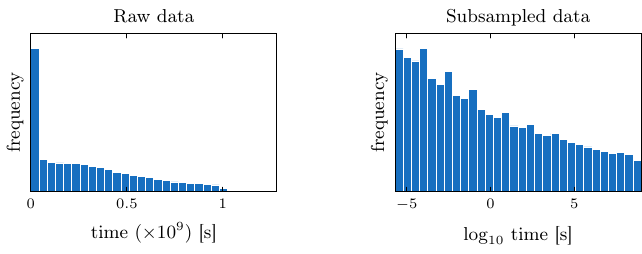}

%% file: fig/input_histograms.tex
\includegraphics{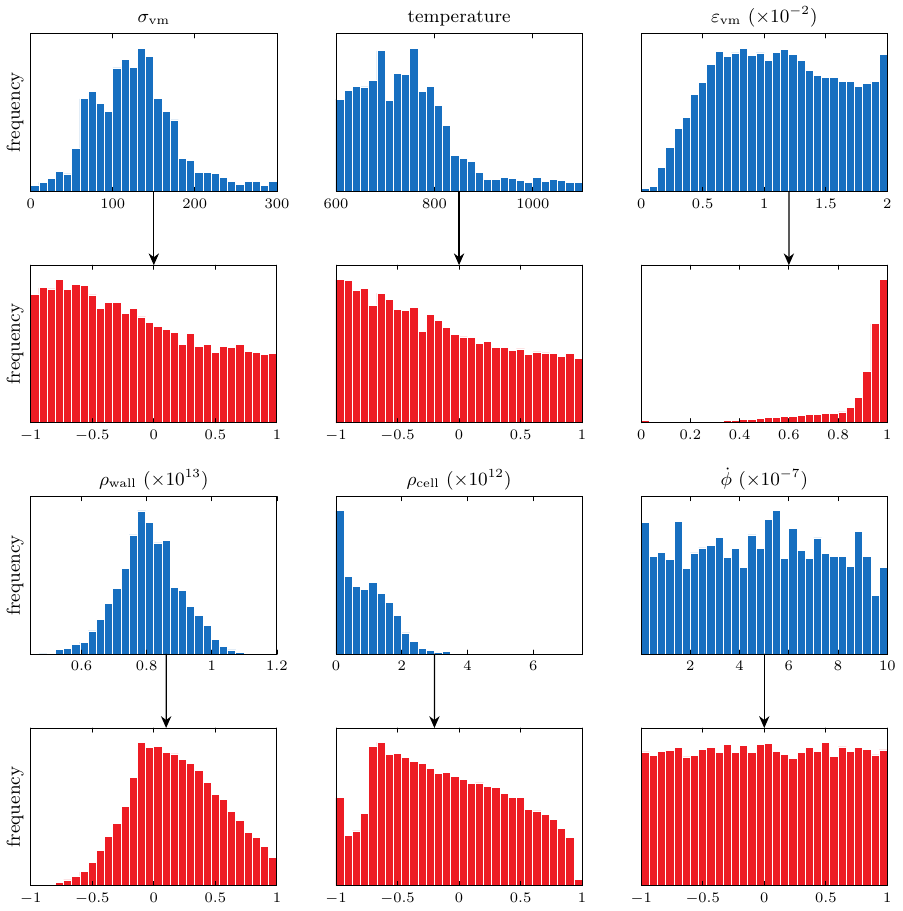}

%% file: fig/output_histograms.tex
\includegraphics{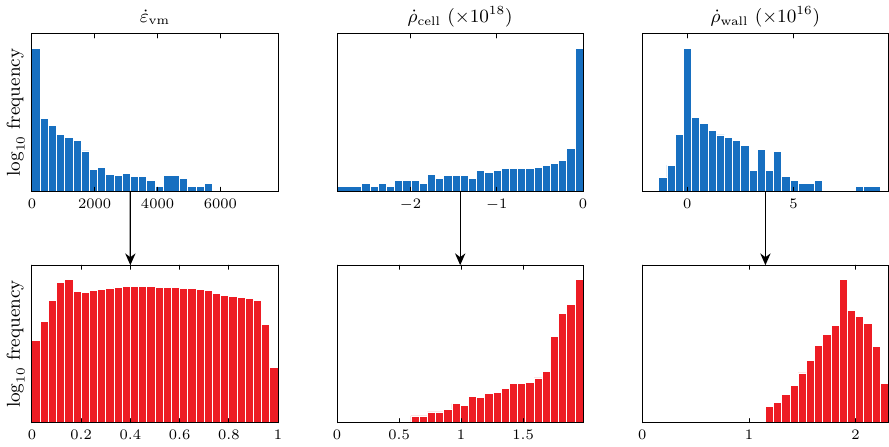}